\documentclass[a4paper,11pt]{article}
\pdfoutput=1
\usepackage{jcappub}
\usepackage{mathtools}
\usepackage{orcidlink}
\newcommand{\Hrh}{H_\text{RH}}
\newcommand{\Trh}{T_\text{RH}}

\newcommand{\arh}{a_\text{RH}}

\title{Freeze-in and ultra-relativistic freeze-out during general reheating scenarios}

\author{Kuldeep Deka \orcidlink{0000-0002-0064-5488}}
\affiliation{New York University Abu Dhabi\\
PO Box 129188, Saadiyat Island, Abu Dhabi, United Arab Emirates}

\emailAdd{kuldeep.deka@nyu.edu}

\abstract{
The dark-matter relic abundance can depend sensitively on the thermal history
before radiation domination. We derive a general analytic framework for
dark-matter production from the Standard Model bath during a non-instantaneous
reheating era, unifying freeze-in, ultra-relativistic freeze-out and the
approach to ordinary non-relativistic freeze-out. The reheating background is
described by an effective equation-of-state parameter $\omega$ and a cooling
index $\alpha$, while the dark-matter interaction rate is parametrised by an
effective scale $\Lambda$ and a leading temperature power $n$. We show that the
production history is organised by two critical temperature exponents: one controls
whether a thermalised relativistic species decouples during reheating or after
radiation domination begins, and the other controls whether post-decoupling
production is infrared dominated, ultraviolet dominated or logarithmic. We
derive analytic relic yields in the main regimes, including both the
entropy-diluted freeze-out contribution and the post-freeze-out production
term. These results explain the scaling of relic-density contours and are
checked against numerical Boltzmann solutions. For matter-like reheating our
framework reproduces the known IR/UV ultra-relativistic freeze-out structure,
while more general reheating histories can shift the same microscopic
interaction between freeze-in, ultra-relativistic freeze-out and ordinary
freeze-out regimes.
}

\begin{document}
\begin{flushright}
\end{flushright}
\maketitle
\flushbottom

\section{Introduction}
\label{sec:intro}

The thermal history of the Universe before Big Bang Nucleosynthesis (BBN) is
only weakly constrained. In the standard relic-density calculation, radiation
domination is often assumed to begin immediately after inflation, so that the
Hubble rate and the plasma temperature follow their radiation-dominated
scalings throughout the era relevant for dark-matter production. This need not
be the case. The energy stored in the inflaton, or in another long-lived
dominant component, can be transferred to the Standard Model (SM) bath over an
extended period. During such a low-temperature reheating (LTR) era, both the
expansion rate and the temperature evolution can differ substantially from the
radiation-dominated behaviour, with important consequences for relic abundances
\cite{Kofman:1997yn,Giudice:2000ex,Allahverdi:2010xz,Amin:2014eta,
Allahverdi:2020bys,Batell:2024dsi}.

Dark matter (DM) production in non-standard cosmologies has been studied in a
variety of settings. Both ordinary thermal freeze-out and freeze-in
\cite{McDonald:2001vt,Hall:2009bx,Bernal:2017kxu} can be modified if radiation
domination begins only after a prolonged reheating stage. Low-temperature
reheating can change the viable parameter space of non-relativistic thermal
relics and can strongly affect freeze-in when the production rate grows with
temperature \cite{Giudice:2000ex,Chung:1998rq,Garcia:2017tuj,Garcia:2018wtq,
Drees:2017iod,Drees:2018dsj,Bernal:2019mhf,Bernal:2020bfj,
Bernal:2020gzm,Barman:2022tzk,Bernal:2024yhu,Bernal:2024ndy,Cosme:2023xpa,
Cosme:2024ndc,Bernal:2025fcl}. In the latter case, the abundance can be
controlled by the largest temperature reached by the bath rather than by the
reheating temperature. This ultraviolet freeze-in behaviour was analysed for
higher-dimensional interactions in Ref.~\cite{Elahi:2014fsa}, and more recently
in general reheating histories parametrised by an equation-of-state parameter
$\omega$ and a temperature-scaling index $\alpha$
\cite{Bernal:2024yhu,Bernal:2025fdr}.

A different possibility is that the DM reaches thermal equilibrium with the SM
bath but decouples while still ultra-relativistic before reheating is completed.
This ultra-relativistic freeze-out (UFO) mechanism is distinct from both
ordinary WIMP freeze-out and pure freeze-in. The final abundance can contain an
entropy-diluted contribution inherited from freeze-out, together with additional
production after decoupling. UFO during non-instantaneous reheating was recently
studied in Refs.~\cite{Henrich:2025sli,Henrich:2025gsd}, where it was shown
that a relativistically decoupled relic can nevertheless be sufficiently cold by
structure formation and can interpolate between WIMP-like and FIMP-like
behaviour. Related work has considered explicit $Z'$-portal realisations and
direct-detection prospects \cite{Henrich:2025pca,Henrich:2026tox}.

The purpose of this paper is to put freeze-in and UFO during reheating in a
single analytic classification. UFO during matter-like reheating and UV
freeze-in during general reheating histories have been studied previously.
Here we ask a complementary question: how does a single microscopic interaction
move between pure freeze-in, ultra-relativistic freeze-out during reheating,
post-reheating relativistic freeze-out and ordinary non-relativistic freeze-out
as the pre-BBN expansion and entropy-production history is varied?

We describe this history by two effective parameters: the equation-of-state
parameter \(\omega\) of the dominant component and the cooling index \(\alpha\)
of the radiation bath, \(T\propto a^{-\alpha}\). Once these are specified, the
temperature dependence of the Hubble rate follows. The resulting production
history is organised by two critical temperature exponents, \(n_c\) and \(n^*\),
which are derived below.

This classification leads to compact analytic expressions for the relic yield
in the main regimes. In the UFO branch we keep both pieces of the final
abundance: the entropy-diluted freeze-out inheritance and the post-freeze-out
production. We also derive the interaction scale required to reproduce the
observed DM abundance. These formulae make the shapes of relic-density contours
transparent: depending on the regime, the relevant physical scale can be
$\Trh$, the freeze-out temperature, the initial temperature of the monotonic
cooling era, or the mass threshold.

We compare the analytic expectations with numerical solutions of the Boltzmann
equation. The numerical treatment uses the full massive Maxwell--Boltzmann
equilibrium density, allowing us to follow the transition from
ultra-relativistic to non-relativistic decoupling. We present relic-density
contours in the $(m_\chi,\Trh)$, $(m_\chi,\Lambda)$ and $(\Lambda,\Trh)$ planes.
Matter-like reheating is used as the reference benchmark, while quartic-like
reheating and kination illustrate how the same interaction can fall into
different production regimes in different reheating backgrounds
\cite{Spokoiny:1993kt,Ferreira:1997hj,Salati:2002md}.

The paper is organised as follows. In Sec.~\ref{sec:reheating} we introduce the
effective reheating parametrisation used throughout. In
Sec.~\ref{sec:conventions} we define the thermodynamic quantities and the DM
interaction rate. Sec.~\ref{sec:boltzmann} presents the Boltzmann equation and
the freeze-out criteria. Sec.~\ref{sec:criticals} introduces the two critical
indices and the associated regimes. Sec.~\ref{sec:relic} gives the analytic
yields and relic-density scalings. Sec.~\ref{sec:bounds} summarises the
consistency conditions used in the numerical scans. Sec.~\ref{sec:examples}
presents representative relic-density contours. We conclude in
Sec.~\ref{sec:conclusions}.

\section{Parametrising cosmic reheating}
\label{sec:reheating}

We use an effective $(\omega,\alpha)$ description of the reheating background,
following the notation of Ref.~\cite{Bernal:2024yhu}. The parameter $\omega$
controls the dilution of the dominant post-inflationary component, while
$\alpha$ controls how the temperature of the radiation bath changes with the
scale factor. This description is not meant to specify a unique microscopic
model of reheating. Rather, it isolates the two background scalings that enter
the DM Boltzmann equation.

\subsection{Background evolution and temperature scaling}

Before radiation domination, we assume that the total energy density is
dominated by a component $\phi$ with averaged equation-of-state parameter
$\omega$. Neglecting the backreaction of the energy transfer on the leading
scaling, its energy density evolves as
\begin{equation}
  \rho_\phi(a)\propto a^{-3(1+\omega)} .
\end{equation}
The radiation bath sourced by this component has energy and entropy densities
\begin{equation}
  \rho_R(T)=\frac{\pi^2}{30}\,g_*(T)T^4,
  \qquad
  s(T)=\frac{2\pi^2}{45}\,g_{*s}(T)T^3 .
  \label{eq:rhoR_s_def}
\end{equation}

We define the reheating temperature $\Trh$ as the temperature at which the
Universe enters radiation domination, and use
\begin{equation}
  \Hrh\equiv H(\Trh)
  =
  \sqrt{\frac{\pi^2 g_{\rm RH}}{90}}\,
  \frac{\Trh^2}{M_P},
  \qquad
  g_{\rm RH}\equiv g_*(\Trh),
  \label{eq:H_RH_explicit}
\end{equation}
where $M_P\simeq2.435\times10^{18}\,{\rm GeV}$ is the reduced Planck mass. An
alternative definition based on $\rho_\phi\simeq\rho_R$ differs only by
order-one factors and does not affect the scaling results below. Throughout the
analytic discussion we use the convention in Eq.~\eqref{eq:H_RH_explicit}.

For analytic work we neglect the mild temperature dependence of $g_*$ and
$g_{*s}$. The Hubble rate is then parametrised as
\begin{equation}
  H(a)=\Hrh\times
  \begin{dcases}
    \left(\dfrac{\arh}{a}\right)^{\frac{3(1+\omega)}{2}},
    & a\le\arh,\\[6pt]
    \left(\dfrac{\arh}{a}\right)^2,
    & a\ge\arh,
  \end{dcases}
  \label{eq:Hubble_a}
\end{equation}
where $a=\arh$ corresponds to $T=\Trh$. The second line is the usual
radiation-dominated scaling.

During the reheating stage we take the temperature to obey
\begin{equation}
  T(a)=\Trh\times
  \begin{dcases}
    \left(\dfrac{\arh}{a}\right)^\alpha,
    & a_I\le a\le\arh,\\[6pt]
    \left(\dfrac{g_{*s}(\Trh)}{g_{*s}(T)}\right)^{1/3}
    \dfrac{\arh}{a},
    & a\ge\arh .
  \end{dcases}
  \label{eq:T_of_a}
\end{equation}
Here $a_I$ denotes the beginning of the monotonic cooling epoch considered in
this work, and $T_I\equiv T(a_I)$ is the corresponding maximum temperature of
the bath. In a full perturbative reheating solution the temperature can first
rise and then fall; our effective description starts after this maximum has
been reached. We focus on $\alpha>0$, so that $T_I>\Trh$. Nearly
constant-temperature trajectories with $\alpha\simeq0$ require a separate
treatment because the map between $T$ and $a$ becomes singular in the strict
$\alpha=0$ limit.

Combining Eqs.~\eqref{eq:Hubble_a} and \eqref{eq:T_of_a}, the Hubble rate during
LTR can be written as
\begin{equation}
  H(T)=\Hrh
  \left(\frac{T}{\Trh}\right)^{\gamma_2},
  \qquad
  \gamma_2\equiv\frac{3(1+\omega)}{2\alpha} .
  \label{eq:H_scaling_LTR}
\end{equation}
This power-law form is the only background input needed for most of the
analytic DM calculation.

Not every pair $(\omega,\alpha)$ describes a reheating stage in which radiation
actually catches up with the dominant component. Since
$\rho_R\propto T^4\propto a^{-4\alpha}$ during the LTR epoch, the radiation
fraction scales as
\begin{equation}
  \frac{\rho_R}{\rho_\phi}
  \propto
  a^{3(1+\omega)-4\alpha} .
\end{equation}
Thus, within the power-law approximation, a viable reheating trajectory requires
\begin{equation}
  3(1+\omega)-4\alpha>0 .
  \label{eq:reheating_viability_condition}
\end{equation}
The marginal case requires a more model-dependent treatment and will not play a
role in the benchmark scans below.

\subsection{Boltzmann equations for reheating}

The energy transfer from $\phi$ to radiation can be described effectively by
\begin{align}
  \frac{d\rho_\phi}{dt}+3(1+\omega)H\rho_\phi
  &=-\Gamma(a)\rho_\phi,
  \label{eq:BErho_phi}
  \\[4pt]
  \frac{d\rho_R}{dt}+4H\rho_R
  &=+\Gamma(a)\rho_\phi,
  \label{eq:BErho_R}
\end{align}
where $\Gamma(a)$ is an effective energy-transfer rate. It may represent
perturbative decays, annihilations, or a more complicated averaged transfer of
energy into the bath. We do not solve Eqs.~\eqref{eq:BErho_phi}--\eqref{eq:BErho_R}
explicitly. Their solutions are used only through the effective scalings
specified above.

For the simple interpretation of Eq.~\eqref{eq:BErho_R} as a positive
source term into radiation, the scaling \(\rho_R\propto a^{-4\alpha}\) implies
\[
   \dot\rho_R+4H\rho_R=4(1-\alpha)H\rho_R ,
\]
so positive energy injection corresponds to \(\alpha<1\), with \(\alpha=1\)
describing adiabatic redshifting of an already produced radiation bath.
Values with \(\alpha>1\), if considered, should be interpreted as an effective
phenomenological scaling rather than as a decay-like positive source of
radiation.

\subsection{Microphysical realisations of \texorpdfstring{$(\omega,\alpha)$}{(omega, alpha)}}

The pair $(\omega,\alpha)$ can arise from different microscopic reheating
mechanisms. For an oscillating scalar in a monomial potential
$V(\phi)\propto\phi^p$, the averaged equation of state is
\cite{Amin:2014eta}
\begin{equation}
  \omega=\frac{p-2}{p+2} .
  \label{eq:w_monomial}
\end{equation}
The cooling index $\alpha$ depends on how the dominant component transfers
energy into radiation. The standard perturbative matter-like benchmark has
$(\omega,\alpha)=(0,3/8)$ \cite{Giudice:2000ex}. Other values occur for
monomial condensates, annihilation-driven reheating, perturbative decays into
bosons or fermions, resonant transfer, and kination-like histories in which the
dominant component redshifts faster than radiation. We collect representative
examples in Appendix~\ref{app:microphysics}. In the main text, however,
$(\omega,\alpha)$ are treated as phenomenological inputs.

\section{Conventions and dark matter interactions}
\label{sec:conventions}

We now specify the interaction between $\chi$ and the SM bath. The radiation
energy and entropy densities were defined in Eq.~\eqref{eq:rhoR_s_def}. In the
analytic expressions we take the relativistic degrees of freedom to be constant
over the temperature range of interest and write
\begin{equation}
  g_{\rm RH}\equiv g_*(\Trh),
  \qquad
  g_{s,{\rm RH}}\equiv g_{*s}(\Trh) .
  \label{eq:gstar_RH_defs}
\end{equation}
The temperature dependence of the $g$-factors can be reinstated in the numerical
solution and affects only order-one normalisations in the analytic scalings.

\subsection{Dark matter interactions and reaction rates}

We assume that $\chi$ is produced from the thermal SM bath through relativistic
$2\leftrightarrow2$ processes. The thermally averaged cross section is written
as
\begin{equation}
  \langle\sigma v\rangle(T)
  =
  c_\sigma\,
  \frac{T^n}{\Lambda^{n+2}} .
  \label{eq:sigmav_param}
\end{equation}
Here $\Lambda$ is an effective interaction scale, $n$ gives the leading
temperature dependence, and $c_\sigma$ is a dimensionless coefficient containing
spin, colour, gauge and Lorentz-structure factors. In the numerical benchmarks
we set $c_\sigma=1$; a different value simply rescales the inferred interaction
scale by $\Lambda\to\Lambda/c_\sigma^{1/(n+2)}$. For a contact operator of
dimension $D$, one generically obtains $n=2D-10$. The numerical examples below
focus on $n\ge0$, appropriate for the higher-dimensional contact interactions
of interest.

Equation~\eqref{eq:sigmav_param} should be understood as the leading scaling
over the temperature interval that dominates production or decoupling. If a
mediator threshold lies between $\Trh$ and $T_I$, the cross section can
interpolate between different powers of $T$ and the analysis should be applied
piecewise. Resonances and finite-width effects are likewise model-dependent and
are not included in the single-power EFT treatment used here.

A separate caveat applies when the solution enters a genuinely
non-relativistic freeze-out regime. The parametrisation in
Eq.~\eqref{eq:sigmav_param} is meant as a relativistic scaling; once
\(T\lesssim m_\chi\), the thermally averaged rate is no longer fixed by the
single exponent \(n\) alone. For a specified operator it can instead depend on
powers of \(m_\chi\), on velocity or helicity suppressions, and on the spin and
Lorentz structure of the interaction. Therefore the analytic FI and UFO
classification derived below should be interpreted as model-independent only
when the production or decoupling epoch is relativistic. In the numerical
regime plots, regions labelled as non-relativistic freeze-out indicate where
the relativistic single-power description ceases to be universal and  their detailed contours are only illustrative.

For relativistic $\chi$, we write
\begin{equation}
  n_{\rm eq}(T)=c_{\rm eq}T^3,
  \qquad
  c_{\rm eq}\equiv \xi_\chi g_\chi\frac{\zeta(3)}{\pi^2},
  \label{eq:neq_rel_def}
\end{equation}
where $g_\chi$ counts the internal degrees of freedom and
$\xi_\chi=1$ for bosons, while $\xi_\chi=3/4$ for fermions. The annihilation
reaction density and the equilibrium interaction rate per particle are then
\begin{align}
  \gamma_a(T)
  &\equiv
  \langle\sigma v\rangle n_{\rm eq}^2
  =
  C_\gamma\,
  \frac{T^{n+6}}{\Lambda^{n+2}},
  \qquad
  C_\gamma\equiv c_\sigma c_{\rm eq}^2,
  \label{eq:gammaa_def}
  \\[4pt]
  \Gamma_{\rm eq}(T)
  &\equiv
  n_{\rm eq}\langle\sigma v\rangle
  =
  C_\Gamma\,
  \frac{T^{\gamma_1}}{\Lambda^{n+2}},
  \qquad
  \gamma_1\equiv n+3,
  \qquad
  C_\Gamma\equiv c_\sigma c_{\rm eq} .
  \label{eq:Gamma_eq_def}
\end{align}
The comparison between the rate exponent $\gamma_1$ and the background exponent
$\gamma_2$ will determine whether decoupling occurs during reheating or after
radiation domination begins.

The relativistic equilibrium yield is
\begin{equation}
  Y_{\rm eq}
  \equiv
  \frac{n_{\rm eq}}{s}
  =
  \frac{45\zeta(3)}{2\pi^4}
  \frac{\xi_\chi g_\chi}{g_{*s}}
  \simeq
  0.278\,
  \frac{\xi_\chi g_\chi}{g_{*s}} .
  \label{eq:Yeq_rel_def}
\end{equation}
It is independent of temperature and of $m_\chi$ as long as $\chi$ remains
relativistic. In the numerical calculation we use the full massive
Maxwell--Boltzmann equilibrium density to follow the transition to
semi-relativistic and non-relativistic freeze-out. This changes only order-one
normalisations in the ultra-relativistic limit, while leaving the regime
classification and contour slopes unchanged.

Throughout the paper the source of $\chi$ is assumed to be scattering from the
thermal SM bath. Direct inflaton decays or annihilations into $\chi$ are not
included. If present, such channels would add a separate source term to the
Boltzmann equation and could modify the freeze-in branch or the initial
condition before thermalisation. In the UFO regime, any earlier abundance is
erased once $\chi$ reaches equilibrium, but late direct production from the
reheating sector would have to be included in a model-dependent analysis.

\section{Boltzmann equation and freeze-out conditions}
\label{sec:boltzmann}

We now write the Boltzmann equation in a form suited to a period of low-temperature
reheating (LTR). The useful variable is the comoving number of \(\chi\), since
the visible-sector comoving entropy is not conserved before reheating is completed.

\subsection{Comoving formulation}

Let \(n_\chi\) denote the physical number density of \(\chi\), and define
\begin{equation}
  N\equiv n_\chi a^3 .
\end{equation}
For \(2\leftrightarrow2\) interactions with the thermal bath,
\begin{equation}
  \frac{dn_\chi}{dt}+3Hn_\chi
  =-
  \langle\sigma v\rangle
  \left(n_\chi^2-n_{\rm eq}^2\right) .
\end{equation}
Equivalently,
\begin{equation}
  \frac{dN}{dt}
  =
  a^3\gamma_a(T)
  \left(1-\frac{N^2}{N_{\rm eq}^2}\right),
  \qquad
  N_{\rm eq}\equiv n_{\rm eq}a^3,
\end{equation}
where \(\gamma_a\equiv \langle\sigma v\rangle n_{\rm eq}^2\). Using
\(dt=da/(Ha)\), this becomes
\begin{equation}
  \frac{dN}{da}
  =
  \frac{a^2}{H(a)}\,\gamma_a(T)
  \left(1-\frac{N^2}{N_{\rm eq}^2}\right).
  \label{eq:boltzmann_full}
\end{equation}
When \(N\ll N_{\rm eq}\), the backreaction term can be neglected and the
same equation reduces to the freeze-in form. When equilibrium is reached,
\(N\) tracks \(N_{\rm eq}\) until the interaction rate falls below the
expansion rate.

\subsection{Thermalisation and decoupling}

The relevant rate-to-Hubble ratio is
\begin{equation}
  R(T)
  \equiv
  \frac{\Gamma_{\rm eq}(T)}{H(T)}
  =
  \frac{C_\Gamma}{\Lambda^{n+2}}\,
  \frac{T^{\gamma_1}}{H(T)} .
  \label{eq:thermalisation_ratio}
\end{equation}
A necessary condition for thermalisation is that \(R(T)\) becomes larger than
unity somewhere along the thermal history. In the analytic classification below
we use
\begin{equation}
  \max_T R(T)\gtrsim 1
\end{equation}
as the practical criterion, with order-one ambiguities absorbed into the
freeze-out constant introduced below. Close to a boundary this instantaneous
criterion should be understood as an analytic proxy for the full Boltzmann
evolution, which is what we use in the numerical scan.

If \(\chi\) thermalises, its decoupling temperature is defined by
\begin{equation}
  R(T_{\rm FO})
  \equiv
  \frac{\Gamma_{\rm eq}(T_{\rm FO})}{H(T_{\rm FO})}
  =c_{\rm fo},
  \qquad
  c_{\rm fo}=\mathcal O(1).
  \label{eq:FO_condition_general}
\end{equation}
The parameter \(c_{\rm fo}\) keeps track of the usual order-one ambiguity in
an analytic freeze-out criterion.

\subsection{Ultra-relativistic freeze-out during LTR}

During LTR the Hubble rate is given by Eq.~\eqref{eq:H_scaling_LTR}. The
freeze-out condition therefore gives
\begin{equation}
  \frac{C_\Gamma}{\Lambda^{n+2}}
  \frac{T_{\rm FO}^{n+3}}
       {\Hrh\left(T_{\rm FO}/\Trh\right)^{\gamma_2}}
  =c_{\rm fo},
\end{equation}
or
\begin{equation}
  T_{\rm FO}^{\gamma_1-\gamma_2}
  =
  \frac{c_{\rm fo}\,\Hrh\,\Lambda^{n+2}}{C_\Gamma}
  \Trh^{-\gamma_2} .
\end{equation}
Using Eq.~\eqref{eq:H_RH_explicit}, the LTR solution is
\begin{equation}
  T_{\rm FO}^{\rm(LTR)}
  =
  \left[
    \frac{c_{\rm fo}}{C_\Gamma}
    \sqrt{\frac{\pi^2 g_{\rm RH}}{90}}\,
    \frac{\Lambda^{n+2}}{M_P}\,
    \Trh^{\,2-\gamma_2}
  \right]^{1/(\gamma_1-\gamma_2)} .
  \label{eq:TFO_LTR}
\end{equation}
This is a physical ultra-relativistic LTR freeze-out only when
\begin{equation}
  \Trh<T_{\rm FO}^{\rm(LTR)}<T_I,
  \qquad
  m_\chi\ll T_{\rm FO}^{\rm(LTR)},
  \qquad
  \gamma_1>\gamma_2 .
  \label{eq:LTR_FO_realisation}
\end{equation}
The last condition ensures that \(R(T)\) decreases as the Universe cools during
LTR. If it is not satisfied, an ultra-relativistic species that remains in
equilibrium through LTR can only decouple later, during radiation domination.

\subsection{Ultra-relativistic freeze-out during RD}

After reheating,
\begin{equation}
  H(T)=
  \sqrt{\frac{\pi^2 g_*}{90}}\,
  \frac{T^2}{M_P} .
\end{equation}
The corresponding relativistic freeze-out temperature satisfies
\begin{equation}
  C_\Gamma\frac{T_{\rm FO}^{n+3}}{\Lambda^{n+2}}
  =
  c_{\rm fo}
  \sqrt{\frac{\pi^2 g_*}{90}}\,
  \frac{T_{\rm FO}^2}{M_P} .
\end{equation}
For \(n>-1\),
\begin{equation}
  T_{\rm FO}^{\rm(RD)}
  =
  \left[
    \frac{c_{\rm fo}}{C_\Gamma}
    \sqrt{\frac{\pi^2 g_*}{90}}\,
    \frac{\Lambda^{n+2}}{M_P}
  \right]^{1/(n+1)} .
  \label{eq:TFO_RD}
\end{equation}
This expression is independent of \(\Trh\), provided that the solution lies
after reheating and remains relativistic,
\begin{equation}
  T_{\rm FO}^{\rm(RD)}<\Trh,
  \qquad
  m_\chi\ll T_{\rm FO}^{\rm(RD)} .
\end{equation}

\section{Critical indices and regimes}
\label{sec:criticals}

The qualitative behaviour of \(\chi\) production during LTR is governed by two
critical powers. The first determines whether relativistic freeze-out can occur
before reheating is completed. The second determines whether FI-like production
is controlled by temperatures near \(\Trh\), by the upper end of the LTR era, or
only logarithmically by the interval between them.

\subsection{First critical index \texorpdfstring{$n_c$}{nc}: LTR versus RD freeze-out}

Using Eq.~\eqref{eq:H_scaling_LTR}, the ratio in
Eq.~\eqref{eq:thermalisation_ratio} can be written as
\begin{equation}
  R(T)
  =
  R_{\rm RH}
  \left(\frac{T}{\Trh}\right)^{\gamma_1-\gamma_2},
  \qquad
  R_{\rm RH}
  \equiv
  \left(\frac{\Gamma_{\rm eq}}{H}\right)_{\Trh}
  =
  \frac{C_\Gamma\Trh^{n+3}}{\Lambda^{n+2}\Hrh} .
  \label{eq:R_scaling}
\end{equation}
Since the temperature decreases during LTR, \(R(T)\) decreases toward reheating
when
\begin{equation}
  \gamma_1-\gamma_2>0
  \quad\Longleftrightarrow\quad
  n>\gamma_2-3 .
\end{equation}
This motivates
\begin{equation}
  \boxed{
  n_c\equiv \gamma_2-3
  =
  \frac{3(1+\omega)}{2\alpha}-3 .
  }
  \label{eq:nc_def}
\end{equation}
For \(n>n_c\), a thermalised \(\chi\) can freeze out during LTR if
\begin{equation}
  R(T_I)>c_{\rm fo}>R_{\rm RH},
\end{equation}
equivalently \(\Trh<T_{\rm FO}<T_I\). If \(R_{\rm RH}>c_{\rm fo}\), equilibrium
survives through reheating and the relativistic freeze-out, if any, occurs in
RD. For \(n\le n_c\), \(R(T)\) does not decrease during LTR, so a distinct LTR
relativistic freeze-out branch is absent.

\subsection{Second critical index \texorpdfstring{$n^*$}{n*}: IR versus UV FI behaviour}

In the FI-like limit \(N\ll N_{\rm eq}\), Eq.~\eqref{eq:boltzmann_full} gives
\begin{equation}
  \frac{dN}{da}
  \simeq
  \frac{a^2}{H(a)}\,\langle\sigma v\rangle n_{\rm eq}^2
  \equiv \mathcal J(a).
  \label{eq:Boltzmann_FIlike}
\end{equation}
Using \(T\propto a^{-\alpha}\), \(H\propto a^{-\alpha\gamma_2}\),
\(n_{\rm eq}\propto T^3\), and \(\langle\sigma v\rangle\propto T^n\), one finds
\begin{equation}
  \mathcal J(a)
  =
  \mathcal J_{\rm RH}
  \left(\frac{a}{\arh}\right)^{\alpha(n^*-n)-1},
  \qquad
  \mathcal J_{\rm RH}
  \equiv
  \frac{\arh^2}{\Hrh}\,
  \langle\sigma v\rangle_{\rm RH}n_{\rm eq}^2(\Trh),
  \label{eq:J_a_scaling}
\end{equation}
where
\begin{equation}
  \boxed{
  n^*
  \equiv
  \gamma_2-6+\frac{3}{\alpha}
  =
  \frac{3(1+\omega)}{2\alpha}-6+\frac{3}{\alpha} .
  }
  \label{eq:n_star_def}
\end{equation}
The integral of \(\mathcal J\) is therefore IR dominated for \(n<n^*\),
logarithmic for \(n=n^*\), and UV dominated for \(n>n^*\). For
\(0<\alpha<1\),
\begin{equation}
  n^*-n_c=\frac{3}{\alpha}-3
  =\frac{3(1-\alpha)}{\alpha}>0 .
\end{equation}
Thus the same interaction can lie in three physically different LTR regimes:
IR-controlled, logarithmic, or UV-controlled. The limit \(\alpha=1\) is special:
then \(n_c=n^*\), the visible comoving entropy is conserved during the LTR-like
era, and the diluted UFO branch discussed below collapses to an undiluted
relativistic relic branch.

\begin{table}[t]
  \centering
  \small
  \begin{tabular}{c c c c}
    \hline
    Regime & Range of \(n\) & Relativistic FO & FI-like LTR integral \\
    \hline
    I
    & \(n\le n_c\)
    & postponed to RD, if thermalised
    & no UFO, IR for pure FI
    \\
    II
    & \(n_c<n<n^*\)
    & LTR UFO possible
    & IR dominated, \(T\sim\Trh\)
    \\
    III
    & \(n=n^*\)
    & LTR UFO possible
    & logarithmic in \(T_{\rm in}/\Trh\)
    \\
    IV
    & \(n>n^*\)
    & LTR UFO possible
    & UV dominated, \(T\sim T_{\rm in}\)
    \\
    \hline
  \end{tabular}
  \caption{Regimes of ultra-relativistic production during LTR for
  \(0<\alpha<1\). Here \(T_{\rm in}=T_I\) for pure freeze-in, while
  \(T_{\rm in}=T_{\rm FO}\) for the FI-like production that occurs after an
  LTR freeze-out.}
  \label{tab:regimes}
\end{table}

\begin{figure}[t]
  \centering
  \includegraphics[width=0.495\linewidth]{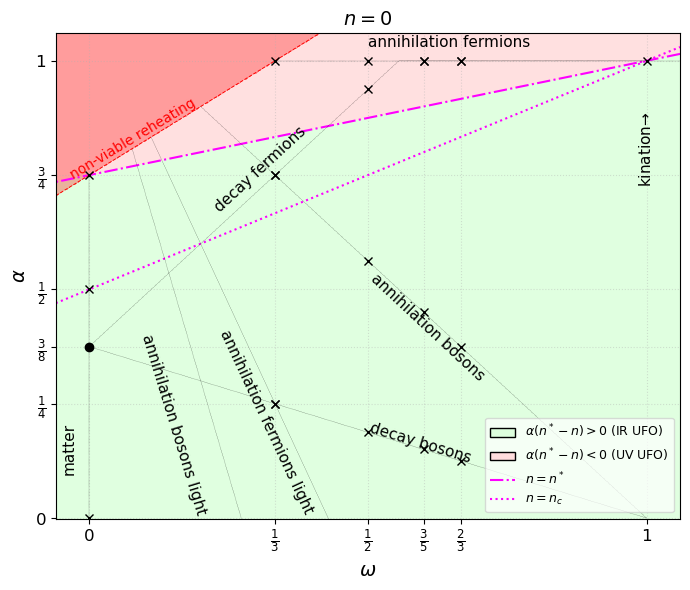}
  \includegraphics[width=0.495\linewidth]{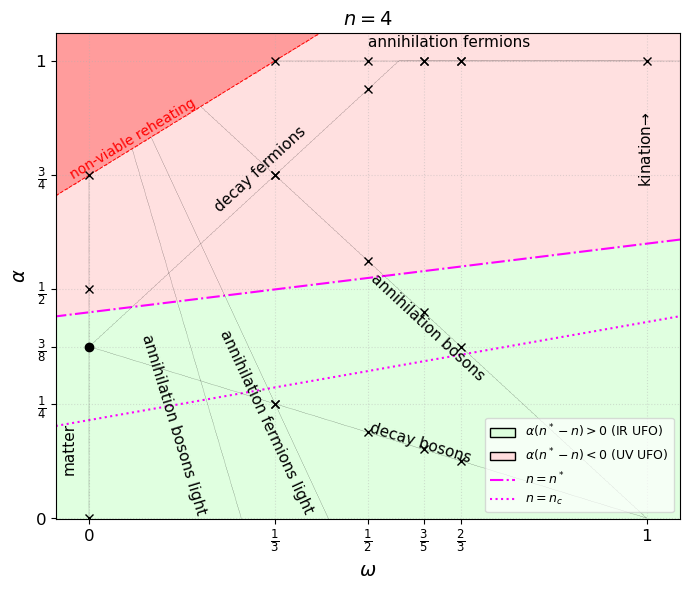}
  \includegraphics[width=0.495\linewidth]{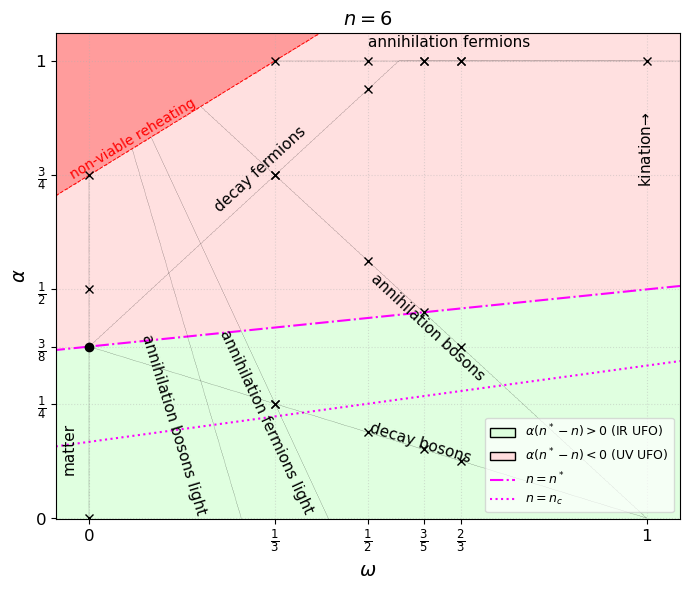}
  \includegraphics[width=0.495\linewidth]{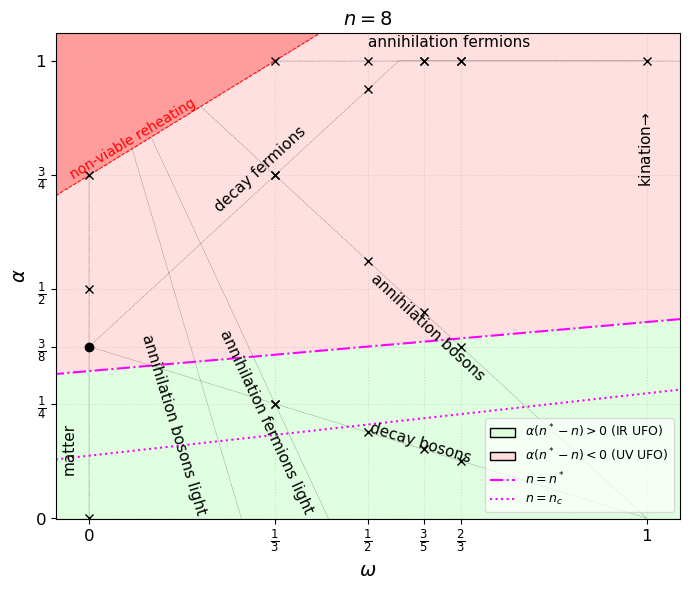}
  \caption{Representative regions in the \((\omega,\alpha)\) plane for several
  interaction powers \(n\). The curves \(n=n_c(\omega,\alpha)\) and
  \(n=n^*(\omega,\alpha)\) separate the freeze-out and FI-like regimes listed
  in Table~\ref{tab:regimes}. The region where the radiation component does not
  overtake the dominant component within the power-law reheating description is
  not used in the phenomenological analysis.}
  \label{fig:alpha_omega}
\end{figure}

\begin{figure}[t]
  \centering
  \includegraphics[width=\linewidth]{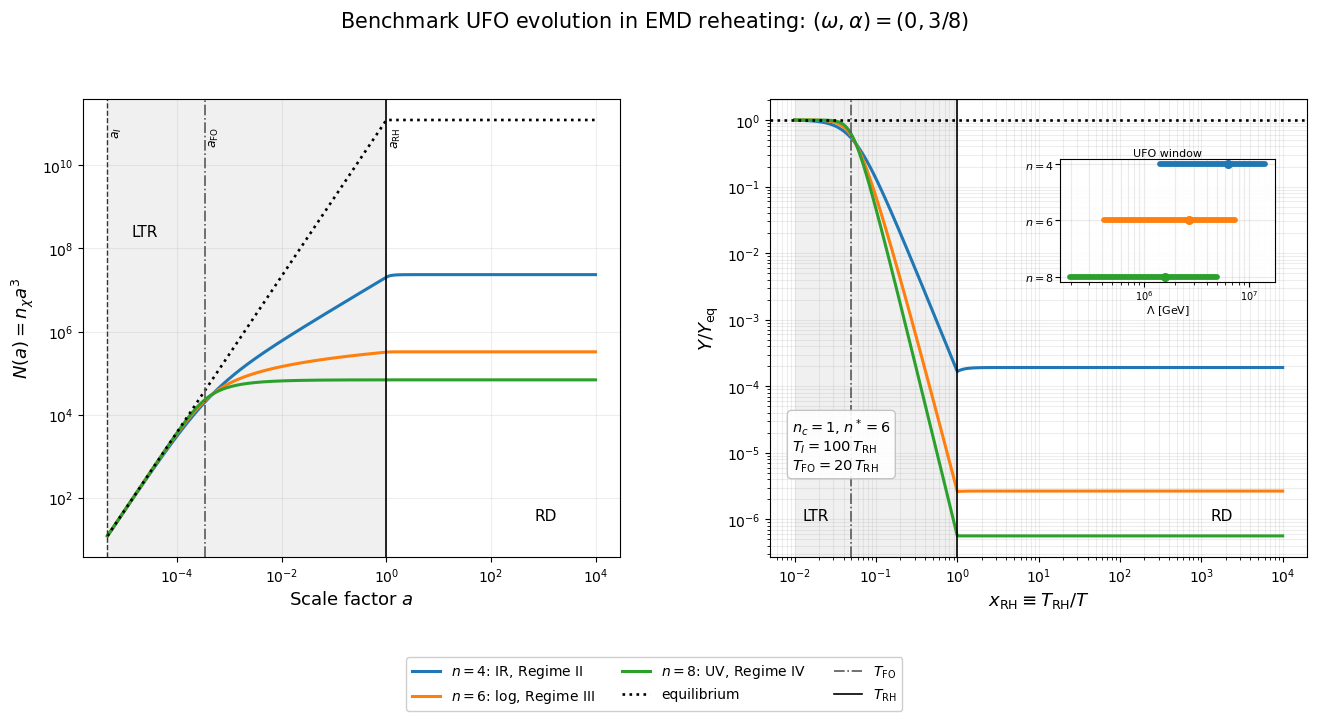}
  \caption{Benchmark evolution for LTR ultra-relativistic freeze-out in the
  matter-like reheating background \((\omega,\alpha)=(0,3/8)\), for which
  \(\gamma_2=4\), \(n_c=1\), and \(n^*=6\). The interaction scale is chosen
  separately for each \(n\) so that all curves decouple at
  \(T_{\rm FO}=20\,\Trh\), with \(T_I=100\,\Trh\). Left: comoving abundance
  \(N=n_\chi a^3\). Right: yield normalised to the relativistic equilibrium
  value as a function of \(x_{\rm RH}\equiv \Trh/T\), so reheating occurs at
  \(x_{\rm RH}=1\). The inset shows the corresponding UFO window in interaction
  scale. The different late-time behaviours reflect whether the post-freeze-out
  source is IR dominated, logarithmic, or UV dominated.}
  \label{fig:UFO_N_sketch}
\end{figure}

\subsection{Interaction window for ultra-relativistic freeze-out during LTR}
\label{subsec:Lambda_window}

For fixed \((\omega,\alpha,\Trh,T_I,n)\), an LTR UFO solution exists only over a
finite interval of interaction strengths. The conditions are
\begin{equation}
  R(\Trh)<c_{\rm fo}<R(T_I),
  \qquad
  n>n_c,
  \qquad
  m_\chi\ll T_{\rm FO} .
\end{equation}
Using Eq.~\eqref{eq:R_scaling}, these become
\begin{align}
  R(\Trh)<c_{\rm fo}
  &\quad\Rightarrow\quad
  \Lambda^{n+2}
  >
  \frac{C_\Gamma\Trh^{n+3}}{c_{\rm fo}\Hrh}
  \equiv
  \Lambda_{\min}^{n+2}(\Trh),
  \label{eq:Lmin_def}
  \\
  R(T_I)>c_{\rm fo}
  &\quad\Rightarrow\quad
  \Lambda^{n+2}
  <
  \frac{C_\Gamma T_I^{n+3}}{c_{\rm fo}H(T_I)}
  \equiv
  \Lambda_{\max}^{n+2}(\Trh,T_I,\omega,\alpha).
  \label{eq:Lmax_def}
\end{align}
Thus
\begin{equation}
  \boxed{
  \Lambda_{\min}<\Lambda<\Lambda_{\max},
  \qquad
  n>n_c .
  }
  \label{eq:Lambda_window_condition}
\end{equation}
For \(\Lambda<\Lambda_{\min}\), the interaction is strong enough to keep
\(\chi\) in equilibrium until reheating, and the relativistic freeze-out branch
is the RD one. For \(\Lambda>\Lambda_{\max}\), the interaction is too weak to
thermalise \(\chi\) during LTR; the abundance is then determined by freeze-in.
Figure~\ref{fig:UFO_N_sketch} illustrates the meaning of this window for a
matter-like reheating benchmark.

\section{Analytic yields and relic interaction scales}
\label{sec:relic}

We now collect the analytic expressions used to interpret the numerical relic-density
contours. The main point is that an LTR UFO abundance contains two pieces: the
equilibrium abundance inherited at decoupling, diluted by the entropy released before
\(\Trh\), and the FI-like production that continues after freeze-out. Which piece sets the
final scaling depends on the position of \(n\) relative to \(n_c\) and \(n^*\).

Throughout this section the final yield is defined after reheating,
\begin{equation}
  Y_0\equiv \frac{N_0}{\mathcal S_{\rm RH}},
  \qquad
  \mathcal S_{\rm RH}\equiv s(\Trh)\arh^3 .
\end{equation}
Neglecting the mild late-time variation of \(g_{*s}\), this is the yield that
enters the present abundance,
\begin{equation}
  \Omega_\chi h^2
  =
  2.742\times10^8
  \left(\frac{m_\chi}{\rm GeV}\right)Y_0 .
\end{equation}
The observed abundance, \(\Omega_{\rm DM}h^2\simeq0.12\)~\cite{Planck:2018vyg},
therefore corresponds to
\begin{equation}
  Y_{\rm DM}
  \equiv
  \frac{\Omega_{\rm DM}h^2}{2.742\times10^8}
  \frac{1}{m_\chi/{\rm GeV}}
  \simeq
  4.37\times10^{-10}
  \left(\frac{\rm GeV}{m_\chi}\right).
  \label{eq:YDM_def}
\end{equation}
In the ultra-relativistic regime the yield itself is independent of \(m_\chi\);
all explicit mass dependence of \(\Lambda_{\rm relic}\) comes from
\(Y_{\rm DM}\propto m_\chi^{-1}\).

\subsection{Common Ingredients}

For constant \(g_{*s}\), the comoving entropy during LTR scales as
\(\mathcal S\propto s a^3\propto T^{3-3/\alpha}\). Hence
\begin{equation}
  \frac{\mathcal S(T_{\rm FO})}{\mathcal S_{\rm RH}}
  =
  \left(\frac{\Trh}{T_{\rm FO}}\right)^d,
  \qquad
  d\equiv\frac{3(1-\alpha)}{\alpha}.
  \label{eq:entropy_ratio}
\end{equation}
The parameter \(d\) measures the entropy dilution between freeze-out and
reheating. It is positive for \(0<\alpha<1\) and vanishes for kination-like
cooling with \(\alpha=1\).

At reheating,
\begin{equation}
  R_{\rm RH}
  =
  C_\Gamma M_P
  \sqrt{\frac{90}{\pi^2 g_{\rm RH}}}
  \frac{\Trh^{n+1}}{\Lambda^{n+2}} .
  \label{eq:R_RH_explicit}
\end{equation}
The FI-like contribution produced between \(T_{\rm in}\) and \(\Trh\), where
\(T_{\rm in}=T_I\) for pure FI and \(T_{\rm in}=T_{\rm FO}\) after an LTR
freeze-out, is
\begin{equation}
\label{eq:YFI_master}
Y_{\rm FI}(T_{\rm in}\to \Trh) =
\begin{cases}
\dfrac{Y_{\rm eq}}{\alpha(n^*-n)} R_{\rm RH}
\left[1-\left(\dfrac{\Trh}{T_{\rm in}}\right)^{n^*-n}\right],
& n\neq n^*, \\[1.2em]
\dfrac{Y_{\rm eq}}{\alpha} R_{\rm RH}
\ln\!\left(\dfrac{T_{\rm in}}{\Trh}\right),
& n=n^* .
\end{cases}
\end{equation}
This single expression contains the IR, logarithmic, and UV limits of LTR
freeze-in. The derivation is given in Appendix~\ref{app:analytic_details}.

\subsection{LTR UFO yield}

If \(\chi\) decouples while relativistic at \(T_{\rm FO}>\Trh\), the abundance
inherited from equilibrium is
\begin{equation}
  Y_{\rm inh}
  =
  Y_{\rm eq}
  \left(\frac{\Trh}{T_{\rm FO}}\right)^d .
  \label{eq:YUV_exact}
\end{equation}
After freeze-out the backreaction term is negligible, so the continued
production from the bath is obtained by setting \(T_{\rm in}=T_{\rm FO}\) in
Eq.~\eqref{eq:YFI_master}:
\begin{equation}
\label{eq:Ypost}
Y_{\rm post} =
\begin{cases}
\dfrac{Y_{\rm eq}}{\alpha(n^*-n)} R_{\rm RH}
\left[1-\left(\dfrac{\Trh}{T_{\rm FO}}\right)^{n^*-n}\right],
& n\neq n^*, \\[1.2em]
\dfrac{Y_{\rm eq}}{\alpha} R_{\rm RH}
\ln\!\left(\dfrac{T_{\rm FO}}{\Trh}\right),
& n=n^* .
\end{cases}
\end{equation}
Thus the LTR freeze-out contribution is
\begin{equation}
  Y_{\rm LTR}^{\rm(FO)}=Y_{\rm inh}+Y_{\rm post} .
  \label{eq:YLTR_FO_total}
\end{equation}
If \(m_\chi<\Trh\), a small radiation-dominated FI tail can also be present:
\begin{equation}
  Y_{\rm RD}
  \simeq
  \frac{Y_{\rm eq}}{n+1}
  \left(\frac{\Gamma_{\rm eq}}{H}\right)_{\Trh}
  \left[1-
    \left(\frac{m_\chi}{\Trh}\right)^{n+1}
  \right],
  \qquad
  (n>-1,\;m_\chi<\Trh).
  \label{eq:Y_RD_tail}
\end{equation}
In the contour plots we solve the full Boltzmann equation, but
Eqs.~\eqref{eq:YUV_exact}--\eqref{eq:Y_RD_tail} explain the observed slopes and
transitions.

\subsection{Asymptotic relic scalings}

The leading ultra-relativistic limits are summarised in
Table~\ref{tab:analytic_scalings}. They are the expressions most useful for
reading the numerical contours.

\begin{table}[t]
  \centering
  \small
  \begin{tabular}{p{0.17\linewidth} p{0.26\linewidth} p{0.15\linewidth} p{0.35\linewidth}}
    \hline
    Branch & Condition & Dominant scale & Leading scaling \\
    \hline
    \\
    RD relativistic relic
    & thermalised through reheating; \(m<T_{\rm FO}<\Trh\)
    & \(T_{\rm FO}^{\rm(RD)}\)
    & \(Y_0\simeq Y_{\rm eq}\) \\
    deep IR-UFO
    & \(n_c<n<n^*\), LTR freeze-out
    & \(\Trh\)
    & \(Y_0\simeq \dfrac{Y_{\rm eq}}{\alpha(n^*-n)}R_{\rm RH}\) \\
    Log FI / log-UFO
    & \(n=n^*\)
    & interval length
    & \(Y_0\simeq \dfrac{Y_{\rm eq}}{\alpha}R_{\rm RH}
       \ln(T_{\rm in}/\Trh)\) \\
    IR FI
    & \(n<n^*\), no thermalisation
    & \(\Trh\)
    & \(Y_0\simeq \dfrac{Y_{\rm eq}}{\alpha(n^*-n)}R_{\rm RH}\) \\   
    UV FI
    & \(n>n^*\), no thermalisation
    & \(T_I\)
    & \(Y_0\simeq \dfrac{Y_{\rm eq}}{\alpha(n-n^*)}R_{\rm RH}
       \left(\dfrac{T_I}{\Trh}\right)^{n-n^*}\) \\
    UV-UFO
    & \(n>n^*\), LTR freeze-out
    & \(T_{\rm FO}\)
    & \(Y_0\simeq Y_{\rm eq}A_{\rm UV}
       \left(\dfrac{\Trh}{T_{\rm FO}}\right)^d\) \\
    Threshold IR
    & \(n<n^*\), \(\Trh\ll m_\chi\ll T_{\rm in}\)
    & \(T\sim m_\chi\)
    & \(\Omega_\chi h^2\propto
       \dfrac{\Trh^{n^*+1}m_\chi^{n+1-n^*}}{\Lambda^{n+2}}\) \\ \\
    \hline
  \end{tabular}
  \caption{Leading analytic limits for the relic abundance. In the log branch,
  \(T_{\rm in}=T_I\) for pure FI and \(T_{\rm in}=T_{\rm FO}\) after LTR
  freeze-out. In the UV-UFO branch
  \(A_{\rm UV}\equiv1+c_{\rm fo}/[\alpha(n-n^*)]\).}
  \label{tab:analytic_scalings}
\end{table}

For \(n\le n_c\), a thermalised relativistic species freezes out after
reheating and
\begin{equation}
  Y_0\simeq Y_{\rm eq} .
\end{equation}
Matching the observed abundance then fixes the usual relativistic thermal relic
mass,
\begin{equation}
  m_\chi^{\rm(thermal)}
  \simeq
  1.6\times10^{-9}~{\rm GeV}\,
  \frac{g_{*s}}{\xi_\chi g_\chi} .
  \label{eq:mchi_thermal}
\end{equation}
There is no distinguished \(\Lambda_{\rm relic}\) on this branch; \(\Lambda\)
only has to be small enough to maintain equilibrium until RD freeze-out.

For deep IR-UFO, \(n_c<n<n^*\), and for IR-FI without thermalisation,
\(n<n^*\), production is controlled by \(T\sim\Trh\). For \(T_I\gg\Trh\) or \(T_{\rm FO}\gg\Trh\), respectively,
\begin{equation}
  Y_0\simeq
  \frac{Y_{\rm eq}C_\Gamma M_P}{\alpha(n^*-n)}
  \sqrt{\frac{90}{\pi^2g_{\rm RH}}}
  \frac{\Trh^{n+1}}{\Lambda^{n+2}} .
  \label{eq:IR_FI_final}
\end{equation}
The relic contour therefore follows
\begin{equation}
  \boxed{
  \Lambda_{\rm relic}^{\rm(IR)}
  =
  \left[
    \frac{Y_{\rm eq}C_\Gamma M_P}{\alpha(n^*-n)Y_{\rm DM}}
    \sqrt{\frac{90}{\pi^2g_{\rm RH}}}
    \Trh^{n+1}
  \right]^{1/(n+2)}
  }
  \label{eq:Lambda_relic_FI}
\end{equation}
and hence \(\Lambda_{\rm relic}\propto(m_\chi\Trh^{n+1})^{1/(n+2)}\).
Near \(T_{\rm FO}\sim\Trh\), the inherited term, the post-freeze-out term and
the RD tail must be retained; this is where the numerical solution is most
useful.

At \(n=n^*\), the same result is replaced by a logarithm,
\begin{equation}
  Y_0\simeq
  \frac{Y_{\rm eq}C_\Gamma M_P}{\alpha}
  \sqrt{\frac{90}{\pi^2g_{\rm RH}}}
  \frac{\Trh^{n^*+1}}{\Lambda^{n^*+2}}
  \ln\frac{T_{\rm in}}{\Trh},
  \label{eq:log_FI_final}
\end{equation}
where \(T_{\rm in}=T_I\) for pure FI and \(T_{\rm in}=T_{\rm FO}\) for
post-freeze-out production. Thus
\begin{equation}
  \boxed{
  \Lambda_{\rm relic}^{\rm(log)}
  =
  \left[
    \frac{Y_{\rm eq}C_\Gamma M_P}{\alpha Y_{\rm DM}}
    \sqrt{\frac{90}{\pi^2g_{\rm RH}}}
    \Trh^{n^*+1}
    \ln\frac{T_{\rm in}}{\Trh}
  \right]^{1/(n^*+2)} .
  }
  \label{eq:Lambda_relic_log_FI}
\end{equation}

For \(n>n^*\), pure freeze-in is UV dominated and controlled by \(T_I\):
\begin{equation}
  Y_0^{\rm(FI)}
  \simeq
  \frac{Y_{\rm eq}C_\Gamma M_P}{\alpha(n-n^*)}
  \sqrt{\frac{90}{\pi^2g_{\rm RH}}}
  \frac{\Trh^{n+1}}{\Lambda^{n+2}}
  \left(\frac{T_I}{\Trh}\right)^{n-n^*} .
  \label{eq:UV_FI_final}
\end{equation}
The corresponding \(\Lambda_{\rm relic}^{\rm(FI,UV)}\) is obtained by setting
this expression equal to \(Y_{\rm DM}\); explicitly it is given in
Appendix~\ref{app:analytic_details}. By contrast, in the UV-UFO branch the
upper scale is not \(T_I\) but the freeze-out temperature itself. For
\(T_{\rm FO}\gg\Trh\) and \(d\neq0\),
\begin{equation}
  Y_0^{\rm(UFO,UV)}
  \simeq
  Y_{\rm eq}A_{\rm UV}
  \left(\frac{\Trh}{T_{\rm FO}}\right)^d,
  \qquad
  A_{\rm UV}
  \equiv
  1+\frac{c_{\rm fo}}{\alpha(n-n^*)} .
  \label{eq:UV_with_subleading}
\end{equation}
Combining this with the LTR freeze-out condition gives
\begin{equation}
  \boxed{
  \Lambda_{\rm relic}^{\rm(UFO,UV)}
  =
  \left[
    \frac{C_\Gamma}{c_{\rm fo}}
    M_P
    \sqrt{\frac{90}{\pi^2g_{\rm RH}}}
    \Trh^{n+1}
  \right]^{1/(n+2)}
  \left(
    \frac{Y_{\rm eq}A_{\rm UV}}{Y_{\rm DM}}
  \right)^{(\gamma_1-\gamma_2)/[d(n+2)]} .
  }
  \label{eq:Lambda_relic_UV}
\end{equation}
Therefore
\begin{equation}
  \Lambda_{\rm relic}^{\rm(UFO,UV)}
  \propto
  \Trh^{(n+1)/(n+2)}
  m_\chi^{(\gamma_1-\gamma_2)/[d(n+2)]}
  =
  \Trh^{(n+1)/(n+2)}
  m_\chi^{(n+3-\gamma_2)/[d(n+2)]} .
  \label{eq:Lambda_UV_scaling}
\end{equation}
This is the distinctive UV-UFO scaling: the relic abundance is set by the
freeze-out scale rather than by the maximum bath temperature. For \(\alpha=1\),
\(d=0\) and the inherited relativistic abundance is not diluted; the deep-UFO
yield approaches an undiluted relativistic relic value instead of generating a
new \(m_\chi\)--\(\Lambda\) scaling.

\subsection{Mass thresholds and the end of the relativistic approximation}

The formulae above assume that the relevant production or decoupling temperatures
are larger than \(m_\chi\). When \(m_\chi\gtrsim\Trh\), the lower end of an
IR-dominated FI-like integral is cut off not by reheating but by the threshold
region \(T\sim m_\chi\). For \(n<n^*\) and
\(\Trh\ll m_\chi\ll T_{\rm in}\), the leading scaling becomes
\begin{equation}
  Y_0^{\rm(IR,thr)}
  \simeq
  \widetilde{\mathcal C}_{\rm thr}
  \frac{M_P\Trh^{n^*+1}}{\Lambda^{n+2}}
  m_\chi^{n-n^*},
  \label{eq:Y_IR_threshold}
\end{equation}
and therefore
\begin{equation}
  \Omega_\chi h^2
  \propto
  \frac{\Trh^{n^*+1}m_\chi^{n+1-n^*}}{\Lambda^{n+2}} .
  \label{eq:Omega_IR_threshold}
\end{equation}
For the matter-like reheating benchmark, \((\omega,\alpha)=(0,3/8)\) and
\(n^*=6\), so
\begin{equation}
  \Omega_\chi h^2
  \propto
  \frac{\Trh^7m_\chi^{n-5}}{\Lambda^{n+2}} .
  \label{eq:benchmark_threshold_scaling}
\end{equation}
This explains the change in slope of IR-dominated contours after they cross
\(m_\chi\simeq\Trh\). The logarithmic case \(n=n^*\) is milder: the mass
threshold replaces the lower logarithmic cutoff, giving
\begin{equation}
  \Omega_\chi h^2
  \propto
  \frac{\Trh^{n^*+1}m_\chi}{\Lambda^{n^*+2}}
  \ln\frac{T_{\rm in}}{m_\chi},
  \qquad
  (\Trh\ll m_\chi\ll T_{\rm in}) .
  \label{eq:Omega_log_threshold}
\end{equation}
A genuine non-relativistic freeze-out is a different limit and occurs only when
\(T_{\rm FO}\lesssim m_\chi\). Then the relic density is governed by the usual
Boltzmann-suppressed freeze-out condition, with the LTR Hubble rate and the
subsequent entropy dilution included if decoupling happens before reheating.
The threshold and non-relativistic estimates are derived in
Appendix~\ref{app:analytic_details}; in the numerical analysis we use the full
Maxwell--Boltzmann equilibrium density across the transition.

\section{Bounds and consistency conditions}
\label{sec:bounds}

Before turning to numerical examples, we summarise the consistency conditions
used when scanning parameter space. The BBN bound is imposed as a hard lower limit on $\Trh$. The Ly-$\alpha$ condition is treated as an approximate free-streaming constraint, while the EFT
and unitarity conditions mark the region where the contact-operator description
is no longer self-contained.

\subsection{BBN bound}

Successful BBN requires radiation domination and an approximately thermal SM bath
before light-element synthesis. Existing BBN and neutrino-thermalisation analyses
place the lower reheating temperature in the few-MeV range
\cite{Kawasaki:2000en,Hannestad:2004px,Sarkar:1995dd,deSalas:2015glj,
Hasegawa:2019jsa}. More recent cosmological analyses find somewhat stronger
limits, depending on the data combination and assumptions
\cite{Barbieri:2025lowrh}. Since the precise bound is mildly model dependent,
we impose the conservative benchmark
\begin{equation}
    \Trh \gtrsim T_{\rm BBN},
    \qquad
    T_{\rm BBN}=4~{\rm MeV}.
\end{equation}
In the plots we shade the region $\Trh<T_{\rm BBN}$ as excluded.
\subsection{Lyman-\texorpdfstring{$\alpha$}{alpha} and free streaming}
\label{subsec:lyman}

Lyman-$\alpha$ forest measurements constrain the suppression of small-scale
matter fluctuations relative to cold DM. These constraints are usually quoted
as a lower bound on the mass of a thermal warm-DM relic,
$m_{\rm WDM}^{\rm th}\gtrsim{\rm few}~{\rm keV}$
\cite{Viel:2013apy,Irsic:2017ixq}. Recent high-redshift analyses find limits
in the same range, for example
$m_{\rm WDM}^{\rm th}>5.7~{\rm keV}$ at $95\%$ C.L. in a baseline analysis,
with somewhat weaker limits under more conservative assumptions about the
thermal history of the intergalactic medium \cite{Irsic:2023fqt}.

A precise recast of these bounds for the present setup would require the full
momentum distribution of $\chi$. We do not attempt such a recast here. Instead,
we use a simple free-streaming estimate, normalised to the commonly quoted
thermal-WDM mass scale, to mark the region where the relic is expected to be too
warm.

If $\chi$ is produced or decouples during LTR at a characteristic temperature
$T_{\rm prod}>\Trh$, its momentum redshifts as $p\propto a^{-1}$ while the SM
bath cools as $T\propto a^{-\alpha}$. The momentum of the decoupled component
relative to the SM temperature at reheating is therefore changed by the factor
\begin{equation}
  {\cal C}_{\rm LTR}(T_{\rm prod})
  \equiv
  \left(
    \frac{T_{\rm prod}}{\Trh}
  \right)^{1-\frac{1}{\alpha}} ,
  \qquad
  T_{\rm prod}>\Trh .
  \label{eq:ltr_momentum_factor}
\end{equation}
For production at or after reheating we set
${\cal C}_{\rm LTR}=1$. Thus entropy-producing reheating histories with
$\alpha<1$ cool an early-decoupled $\chi$ population relative to the SM bath,
whereas kination-like histories with $\alpha=1$ do not generate such an
additional cooling factor.

We then impose the approximate free-streaming condition
\begin{equation}
  m_\chi
  \gtrsim
  m_{\rm fs}^{\rm ref}\,
  \eta_{\rm fs}\,
  {\cal C}_{\rm LTR}(T_{\rm prod}) .
  \label{eq:lyman_bound_general}
\end{equation}
Here $m_{\rm fs}^{\rm ref}$ is chosen to be of order the thermal-WDM
Ly-$\alpha$ bound, while $\eta_{\rm fs}$ absorbs order-one information about the
spectrum, the mapping to a thermal-WDM transfer function, and SM entropy release
after reheating. In the numerical plots we use
\begin{equation}
  m_{\rm fs}^{\rm ref}=5~{\rm keV},
  \qquad
  \eta_{\rm fs}=1 .
  \label{eq:lyman_plot_choice}
\end{equation}
The shaded Ly-$\alpha$ region in the figures should therefore be read as an
approximate free-streaming constraint, not as a model-independent exclusion.
The characteristic scale $T_{\rm prod}$ is taken to be the scale controlling the
final abundance.

\subsection{Dark-radiation contribution}
\label{subsec:dark_radiation}

If $\chi$ is still relativistic during BBN, its energy density contributes to
the expansion rate and can be expressed as an effective contribution to
$N_{\rm eff}$. Current CMB measurements are consistent with the Standard Model
radiation density; for example, Planck+BAO gives
$N_{\rm eff}=2.99\pm0.17$, while ACT DR6 finds
$N_{\rm eff}=2.86\pm0.13$ for free-streaming relativistic species
\cite{Planck:2018vyg,AtacamaCosmologyTelescope:2025nti}.

For a relativistic relic with yield $Y_\chi$, the energy density at BBN is
approximately
\begin{equation}
  \rho_\chi
  \simeq
  Y_\chi s_{\rm SM}\,\langle p\rangle .
\end{equation}
Writing
\begin{equation}
  \xi_\chi^{\rm BBN}
  \equiv
  \frac{\langle p\rangle_{\rm BBN}}{T_{\rm BBN}},
\end{equation}
and taking $T_\nu\simeq T_\gamma$ before $e^\pm$ annihilation, one obtains
\begin{equation}
  \Delta N_{\rm eff}^{\rm BBN}
  \simeq
  \frac{16}{21}\,
  g_{*s}^{\rm BBN}\,
  Y_\chi\,
  \xi_\chi^{\rm BBN}.
  \label{eq:DeltaNeff_yield}
\end{equation}
The same LTR momentum factor in Eq.~\eqref{eq:ltr_momentum_factor} enters
$\xi_\chi^{\rm BBN}$, up to order-one spectral and entropy-release factors.

On the relic-density contour,
\begin{equation}
  Y_\chi
  =
  Y_{\rm DM}
  \simeq
  4.4\times10^{-10}
  \left(\frac{\Omega_\chi h^2}{0.12}\right)
  \left(\frac{\rm GeV}{m_\chi}\right),
\end{equation}
and therefore
\begin{equation}
  \Delta N_{\rm eff}^{\rm BBN}
  \simeq
  3.6\times10^{-9}
  \left(\frac{g_{*s}^{\rm BBN}}{10.75}\right)
  \left(\frac{\Omega_\chi h^2}{0.12}\right)
  \left(\frac{\xi_\chi^{\rm BBN}}{1}\right)
  \left(\frac{\rm GeV}{m_\chi}\right).
  \label{eq:DeltaNeff_relic_contour}
\end{equation}
Thus even the illustrative requirement
$\Delta N_{\rm eff}\lesssim0.3$ corresponds to
\begin{equation}
  m_\chi
  \gtrsim
  12~{\rm eV}\,
  \left(\frac{\Delta N_{\rm eff}^{\rm max}}{0.3}\right)^{-1}
  \left(\frac{g_{*s}^{\rm BBN}}{10.75}\right)
  \left(\frac{\Omega_\chi h^2}{0.12}\right)
  \left(\frac{\xi_\chi^{\rm BBN}}{1}\right).
  \label{eq:DeltaNeff_mass_bound}
\end{equation}
For an uncooled relativistic spectrum, $\xi_\chi^{\rm BBN}\simeq3.15$, this is
only a tens-of-eV requirement. This is well below the keV-scale free-streaming
constraint used in Eq.~\eqref{eq:lyman_bound_general}. For the entropy-producing
LTR cases with $\alpha<1$, early decoupling further reduces
$\xi_\chi^{\rm BBN}$ relative to the SM bath. Consequently, for the relic-density
contours shown in this work, the dark-radiation bound is weaker than the
Ly-$\alpha$ free-streaming estimate. We therefore do not show a separate
$\Delta N_{\rm eff}$ shaded region in the plots.

\subsection{Inflationary scale and the maximum temperature \texorpdfstring{$T_I$}{TI}}

In the effective description, the LTR epoch begins at a maximum bath
temperature $T_I$. We parameterise this temperature in terms of an initial
Hubble scale $H_I$ using
\begin{equation}
  H(T)
  =
  \Hrh
  \left(
    \frac{T}{\Trh}
  \right)^{\gamma_2},
  \qquad
  \gamma_2=\frac{3(1+\omega)}{2\alpha}.
\end{equation}
Inverting gives
\begin{equation}
  T_I
  =
  \Trh
  \left(
    \frac{H_I}{\Hrh}
  \right)^{1/\gamma_2}
  =
  \Trh
  \left(
    \frac{H_I}{\Hrh}
  \right)^{\frac{2\alpha}{3(1+\omega)}}.
  \label{eq:TI_from_HI}
\end{equation}
The parameter $H_I$ should not exceed the inflationary Hubble scale, which is
bounded by CMB limits on primordial tensor modes. In slow-roll inflation,
\begin{equation}
  H_*
  =
  \pi M_P
  \sqrt{\frac{A_s r}{2}},
\end{equation}
so bounds on the tensor-to-scalar ratio of order
$r\lesssim{\cal O}(10^{-2})$ imply
$H_*\lesssim{\rm few}\times10^{13}~{\rm GeV}$~\cite{BICEP:2021xfz}. We use
\begin{equation}
  H_I=10^{13}~{\rm GeV}
\end{equation}
as a representative benchmark. Varying $H_I$ mainly affects UV-dominated
regions, where the production integrals are sensitive to the upper limit
$T_I$.

\subsection{EFT validity and a unitarity bound}

The interaction in Eq.~\eqref{eq:sigmav_param} is described by a contact
operator suppressed by the scale $\Lambda$, which should be interpreted as a
proxy for the mass scale of a UV completion. EFT validity requires the typical
centre-of-mass energy of the processes relevant for production or
thermalisation to lie below the cutoff,
\begin{equation}
  \sqrt{s}\sim{\cal O}(T)\lesssim \Lambda .
\end{equation}
In addition, partial-wave unitarity for relativistic $2\to2$ scattering
requires the cross section not to grow without bound. Parametrically,
$\sigma\lesssim 4\pi/s$, which for
$\langle\sigma v\rangle\simeq T^n/\Lambda^{n+2}$ gives the nominal unitarity
temperature
\begin{equation}
  T_{\rm uni}
  =
  (4\pi)^{1/(n+2)}\Lambda .
\end{equation}
We therefore impose the conservative cutoff criterion
\begin{equation}
  T \lesssim T_{\rm cut},
  \qquad
  T_{\rm cut}\equiv \min\{\Lambda,T_{\rm uni}\}.
  \label{eq:Tcut_def}
\end{equation}

For the cases with $n\ge0$ considered in the numerical examples,
$T_{\rm uni}\gtrsim\Lambda$, so the EFT requirement is usually the limiting
one.

\subsection{Characteristic temperatures and EFT/uni\-tarity check}

To implement Eq.~\eqref{eq:Tcut_def} in a compact way, we define a
characteristic temperature \(T_{\rm char}\) at which the contact-operator
description is most stressed for the abundance calculation. The choice of
\(T_{\rm char}\) follows the temperature range that controls the final yield.
This is weaker than requiring the contact operator to describe every part of
the earlier thermalisation history.

For the plots we use the following choices:
\begin{itemize}
  \item \textbf{Pure LTR FI:}
  \begin{itemize}
    \item IR dominated or logarithmic, \(n\le n^*\):
    \[
      T_{\rm char}\simeq \max(\Trh,m_\chi).
    \]
    \item UV dominated, \(n>n^*\):
    \[
      T_{\rm char}\simeq \max(T_I,\Trh,m_\chi).
    \]
  \end{itemize}
  The exactly logarithmic case \(n=n^*\) has no unique endpoint dominance;
  in the plots it is grouped with the lower-end choice. Using the upper end
  \(T_I\) instead would give a more restrictive EFT shading only on this
  boundary case.

  \item \textbf{LTR UFO:}
  \begin{itemize}
    \item IR-dominated post-freeze-out production, \(n<n^*\):
    \[
      T_{\rm char}\simeq \max(\Trh,m_\chi).
    \]
    \item Logarithmic post-freeze-out production, \(n=n^*\):
    \[
      T_{\rm char}\simeq \max(\Trh,T_{\rm FO},m_\chi).
    \]
    \item UV-dominated post-freeze-out production, \(n>n^*\):
    \[
      T_{\rm char}\simeq \max(T_{\rm FO},m_\chi).
    \]
  \end{itemize}
  In the IR-UFO case the final yield is controlled near reheating, even though
  the decoupling event itself occurs earlier at \(T_{\rm FO}>\Trh\).

  \item \textbf{RD or non-relativistic freeze-out, including NR-FO-LTR:}
  \[
    T_{\rm char}\simeq \max(\Trh,T_{\rm FO},m_\chi).
  \]
\end{itemize}
In the figures, a point is marked as passing the EFT/uni\-tarity check when
\begin{equation}
  T_{\rm char}\le T_{\rm cut}.
  \label{eq:EFT_char_check}
\end{equation}
Points violating this condition are shaded as EFT/uni\-tarity unsafe. This
shading should be read as a statement about the single-power contact-operator
description used for the abundance calculation. It does not rule out the
possibility that a specified UV completion gives a consistent description in
the same region.

\subsection{Numerical regime classification}
\label{subsec:classification}

In the numerical scans, each point in a given parameter plane is classified by
solving the Boltzmann equation for the comoving number
$N=a^3n_\chi$ in the matched LTR+RD background, using the full
Maxwell--Boltzmann equilibrium density. When using the rate criterion, $T_{\rm FO}$ is identified with the last downward crossing of $R(T)=\Gamma_{\rm eq}/H=1$ after the species has reached equilibrium. We then define
\begin{equation}
  x_{\rm FO}
  \equiv
  \frac{m_\chi}{T_{\rm FO}} .
\end{equation}
The qualitative regimes used in the plots are:
\begin{itemize}
  \item \textbf{FI}: $\chi$ never reaches equilibrium over the relevant thermal
  history, equivalently $\max_T R(T)<1$ to good approximation.

  \item \textbf{UFO-LTR}: freeze-out occurs during LTR,
  $T_{\rm FO}>\Trh$, and is relativistic, $x_{\rm FO}<1$.

  \item \textbf{NR-FO-LTR}: freeze-out occurs during LTR with
  $x_{\rm FO}\ge1$.

  \item \textbf{FO-RD-rel}: freeze-out occurs after reheating,
  $T_{\rm FO}<\Trh$, and is relativistic, $x_{\rm FO}<1$.

  \item \textbf{WIMP-RD}: freeze-out occurs after reheating with
  $x_{\rm FO}\ge1$.
\end{itemize}
The boundary $x_{\rm FO}=1$ should be understood as a practical convention for
separating relativistic and non-relativistic behaviour; the transition is
smooth in the full numerical solution. The non-relativistic labels are included to show where the relativistic FI/UFO
asymptotics no longer apply. The numerical evolution remains well defined
within the phenomenological single-power ansatz of Eq.~\eqref{eq:sigmav_param},
but a contact-operator interpretation of those regions requires specifying the
non-relativistic matrix element. The same classification is used in all
parameter planes, including $(\Trh,m_\chi)$, $(\Lambda,m_\chi)$ and
$(\Lambda,\Trh)$. We then overlay the relic-density contour
$\Omega_\chi h^2=0.12$ and apply the BBN, Ly-$\alpha$ and EFT diagnostics
described above.

\subsection{Other model-dependent constraints}

Additional constraints can arise once the UV completion of the contact operator
is specified. Depending on the Lorentz structure, mediator mass, and SM charges,
collider searches, direct detection, indirect detection, or cosmological limits
on late-time energy injection may apply. Since our goal is to isolate the
early-Universe production mechanism in an EFT language, we do not impose such
model-dependent bounds. Instead, we use the EFT/unitarity check to indicate where the single-power
contact-operator description is self-contained for the abundance calculation.

\section{Relic-density contours in representative parameter planes}
\label{sec:examples}

We now apply the analytic scalings of Sec.~\ref{sec:relic} to the numerical
$\Omega_\chi h^2=0.12$ contours. The background colours show the regime
classification defined in Sec.~\ref{subsec:classification}, while the solid
curves denote the relic-density contour. Shaded regions mark the BBN bound,
the approximate free-streaming constraint, and the region where the contact
operator is not self-contained.

\subsection{How to read the contour plots}
\label{subsec:plot_reading}

The shapes of the relic-density contours follow from a small number of
asymptotic laws.

First, if $\chi$ remains thermalised through reheating and freezes out in RD
while still relativistic, then
\begin{equation}
  Y_0\simeq Y_{\rm eq},
  \qquad
  \Omega_\chi h^2\propto m_\chi .
  \label{eq:plotscaling_RDrel}
\end{equation}
The relic condition therefore fixes the familiar relativistic thermal relic
mass, $m_\chi\simeq m_\chi^{\rm(thermal)}$, independently of
$\Trh$ and $\Lambda$ within the thermalisation window.

Second, in the IR-dominated ultra-relativistic regime, $n<n^*$ and
$m_\chi\ll\Trh$, both pure IR FI and deep IR-UFO obey
\begin{equation}
  \Omega_\chi h^2
  \propto
  \frac{m_\chi\,\Trh^{n+1}}{\Lambda^{n+2}}.
  \label{eq:plotscaling_IR_UR}
\end{equation}
This is the scaling behind Eq.~\eqref{eq:Lambda_FI_scaling}. In this regime,
deep IR-UFO reduces smoothly to pure IR FI because the inherited freeze-out
abundance is diluted and the post-freeze-out production is controlled near
$\Trh$.

Third, for genuinely IR-dominated branches, $n<n^*$, crossing into the regime
$\Trh\lesssim m_\chi\ll T_{\rm in}$ changes the lower end of the production
integral. The abundance is then controlled by the threshold region
$T\sim m_\chi$, giving
\begin{equation}
  \Omega_\chi h^2
  \propto
  \frac{\Trh^{n^*+1}\,m_\chi^{n+1-n^*}}{\Lambda^{n+2}},
  \qquad
  (n<n^*,\;\Trh\lesssim m_\chi\ll T_{\rm in}).
  \label{eq:plotscaling_IR_thr}
\end{equation}
This threshold-controlled law changes the power-law slope of IR-dominated
contours across the line $m_\chi\simeq\Trh$.

At the logarithmic boundary, $n=n^*$, the mass threshold does not generate a
new power law. Instead, it replaces the lower cutoff of the logarithm,
\begin{equation}
  \ln\frac{T_{\rm in}}{\Trh}
  \;\longrightarrow\;
  \ln\frac{T_{\rm in}}{\max(\Trh,m_\chi)} ,
  \label{eq:log_threshold_cutoff}
\end{equation}
up to order-one threshold corrections. Thus the $n=n^*$ contour shows only mild
logarithmic curvature when $m_\chi$ crosses $\Trh$; the visible turnover occurs
when $m_\chi$ approaches the actual production scale $T_{\rm in}$, typically
$T_{\rm FO}$ for UFO.

Fourth, for UV-dominated pure FI, $n>n^*$, the yield is controlled by the upper
temperature endpoint $T_I$. From Eq.~\eqref{eq:Lambda_relic_UV_FI}, at fixed
$(\Trh,T_I)$ one obtains
\begin{equation}
  \Lambda_{\rm relic}^{\rm(FI,UV)}
  \propto
  m_\chi^{1/(n+2)}.
  \label{eq:plotscaling_UV_FI}
\end{equation}
By contrast, in the UV-dominated UFO branch with $d\neq0$, the abundance is
controlled by the decoupling scale $T_{\rm FO}$. From
Eq.~\eqref{eq:Lambda_UV_scaling},
\begin{equation}
  \Lambda_{\rm relic}^{\rm(UFO,UV)}
  \propto
  \Trh^{\frac{n+1}{n+2}}\,
  m_\chi^{\frac{n+3-\gamma_2}{d(n+2)}},
  \qquad
  d=\frac{3(1-\alpha)}{\alpha}.
  \label{eq:plotscaling_UV}
\end{equation}
This difference between UV-FI and UV-UFO is the origin of the visible
green--blue slope change in the non-EMD UV-dominated examples below.

Finally, the genuinely non-relativistic regions are treated numerically. The
NR-FO-LTR and WIMP-RD regions are shown in the background classification for
completeness, but we do not assign universal analytic slopes to the deep NR
contours. In the benchmarks shown below, these regions either interpolate
between the UR asymptotic branches or lie largely in EFT/uni\-tarity-unsafe
territory.

\subsection{Matter-like reheating benchmark}
\label{subsec:emd_results}

We first consider the standard matter-like reheating benchmark
\begin{equation}
  (\omega,\alpha)=\left(0,\frac38\right),
  \qquad
  \gamma_2=4,
  \qquad
  n_c=1,
  \qquad
  n^*=6,
  \qquad
  d=5.
  \label{eq:emd_benchmark_values}
\end{equation}
The panels below use $n=0,4,6,8$, which respectively probe Regime I,
the IR-dominated branch, the logarithmic boundary, and the UV-dominated branch.

For this benchmark, the threshold-controlled scaling
Eq.~\eqref{eq:plotscaling_IR_thr} reduces to
\begin{equation}
  \Omega_\chi h^2
  \propto
  \frac{\Trh^7 m_\chi^{n-5}}{\Lambda^{n+2}}.
  \label{eq:plotscaling_IR_thr_benchmark}
\end{equation}
This relation will be useful for interpreting the contour bends across
$m_\chi\simeq\Trh$.


\begin{figure}[t]
  \centering
  \includegraphics[width=0.495\linewidth]{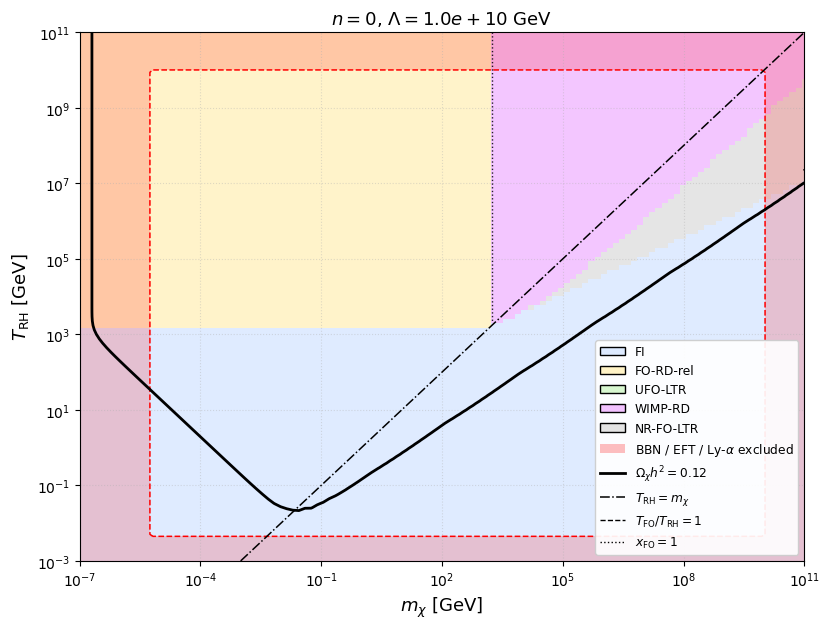}
  \includegraphics[width=0.495\linewidth]{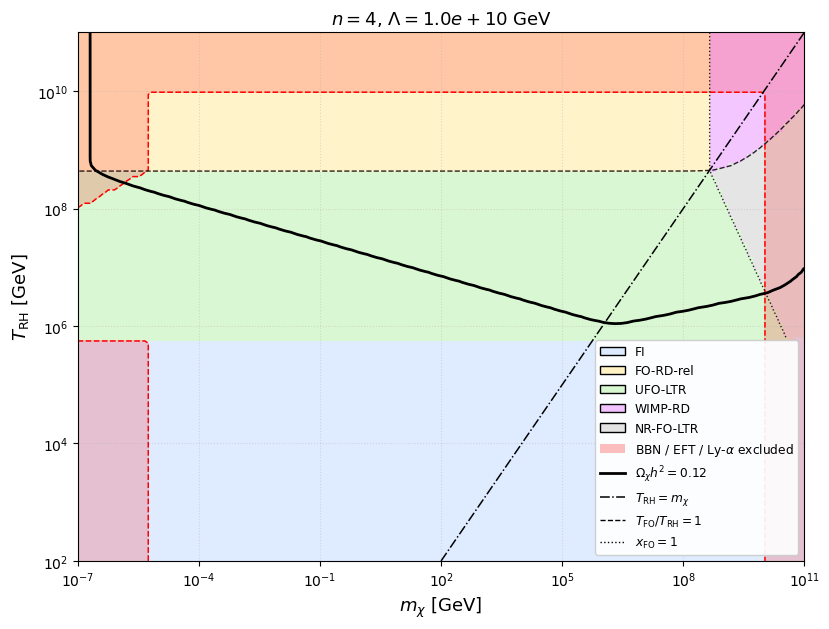}\\[4pt]
  \includegraphics[width=0.495\linewidth]{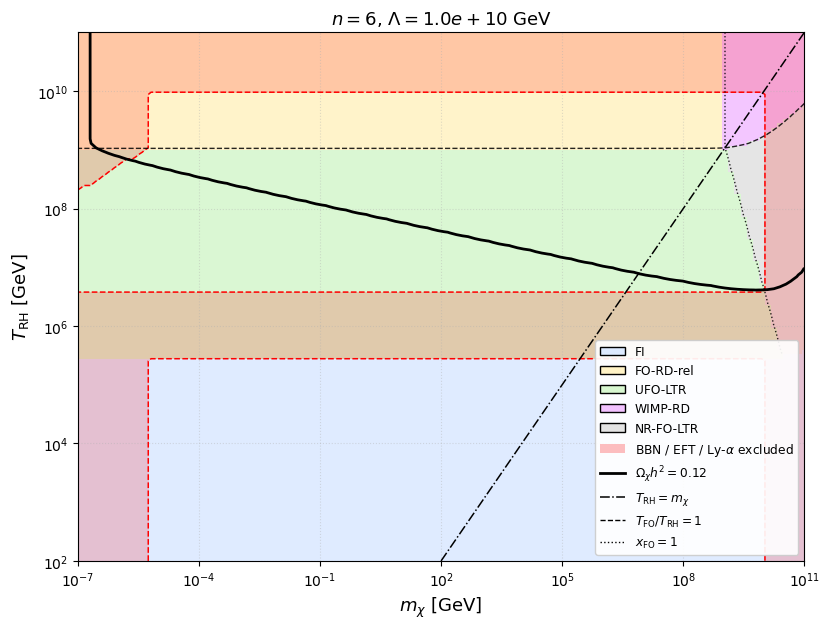}
  \includegraphics[width=0.495\linewidth]{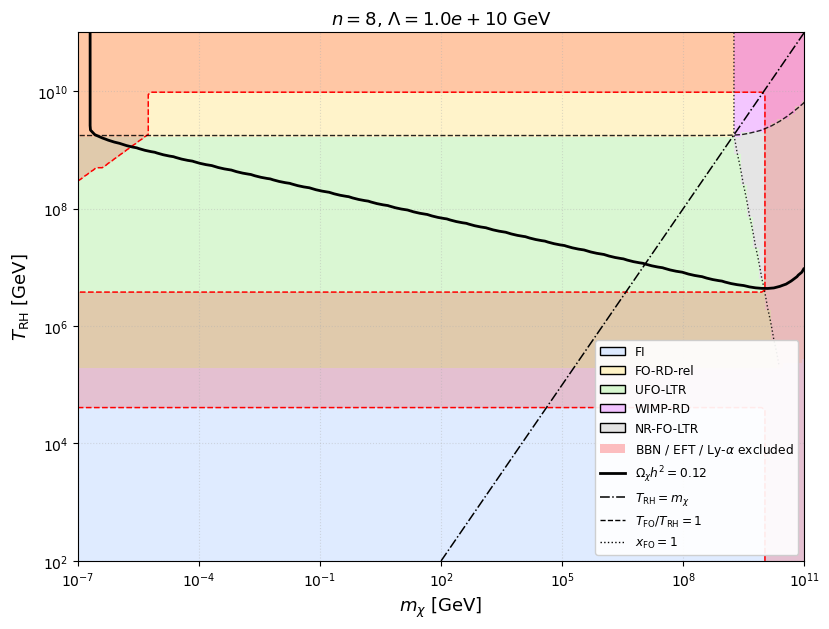}
  \caption{$(m_\chi,\Trh)$ planes for $n=0,4,6,8$, with
  $\Lambda=10^{10}$~GeV and $(\omega,\alpha)=(0,3/8)$. The background colours
  denote the numerical regime classification. The solid curve is the numerical
  $\Omega_\chi h^2=0.12$ contour. Shaded regions are excluded or flagged by the
  consistency conditions of Sec.~\ref{sec:bounds}.}
  \label{fig:Trh_vs_m}
\end{figure}

\paragraph{\boldmath The $(m_\chi,\Trh)$ plane.}

Figure~\ref{fig:Trh_vs_m} shows how the relic contour moves as the reheating
temperature is varied at fixed $\Lambda$. Since
\begin{equation}
  R_{\rm RH}\propto \frac{\Trh^{n+1}}{\Lambda^{n+2}},
\end{equation}
increasing $\Trh$ strengthens the interaction at reheating. The system therefore
moves from weakly coupled FI, through LTR freeze-out, and eventually to
freeze-out after reheating.

The low-mass RD-relativistic branch is universal. Whenever $\chi$ remains in
equilibrium through reheating and freezes out in RD while still relativistic,
Eq.~\eqref{eq:plotscaling_RDrel} fixes
$m_\chi\simeq m_\chi^{\rm(thermal)}$. Since the horizontal axis is $m_\chi$,
this appears as an approximately vertical contour segment. In the $n=0$ panel,
$n<n_c$, so relativistic LTR freeze-out is absent and this is the only thermal
freeze-out branch.

For $n=4$, one has $n_c<n<n^*$, so the LTR branch is IR dominated. In the UR
domain $m_\chi\ll\Trh$, Eq.~\eqref{eq:plotscaling_IR_UR} gives
\begin{equation}
  \Trh\propto m_\chi^{-1/(n+1)}.
  \label{eq:plotslope_TRHm_IRUR}
\end{equation}
Once the contour crosses into $\Trh\lesssim m_\chi$, the threshold-controlled
law in Eq.~\eqref{eq:plotscaling_IR_thr} takes over,
\begin{equation}
  \Trh
  \propto
  m_\chi^{-\frac{n+1-n^*}{n^*+1}}.
  \label{eq:plotslope_TRHm_IRthr}
\end{equation}
For the EMD benchmark this becomes
\begin{equation}
  \Trh\propto m_\chi^{(5-n)/7}.
  \label{eq:plotslope_TRHm_IRthr_benchmark}
\end{equation}
Thus for $n=4$ the IR branch changes from a negative slope above the line
$m_\chi=\Trh$ to a positive slope below it.

The logarithmic boundary, illustrated by the $n=6$ panel, behaves differently
from a genuinely IR-dominated contour. Since $n=n^*$, crossing the line
$m_\chi=\Trh$ only changes the lower cutoff of the logarithm,
as in Eq.~\eqref{eq:log_threshold_cutoff}. For EMD this gives
\begin{equation}
  \Omega_\chi h^2
  \propto
  \frac{m_\chi\,\Trh^7}{\Lambda^8}
  \ln\frac{T_{\rm in}}{\max(\Trh,m_\chi)} ,
\end{equation}
up to order-one threshold corrections. The leading power-law slope is therefore
almost unchanged across $m_\chi=\Trh$. The visible turnover occurs later, when
$m_\chi$ approaches the actual production scale, typically
$T_{\rm FO}$ in the UFO branch. This is why the $n=6$ panel resembles the
UV-dominated case more than the IR $n=4$ case with respect to the location of the turn.

For $n=8$, the production is UV dominated. In this case the relic is controlled
by the decoupling scale $T_{\rm FO}$ rather than directly by $\Trh$, and the
contour is not tied to the line $m_\chi=\Trh$. Instead, the visible turnover
occurs when $m_\chi$ approaches the true production or decoupling scale.


\begin{figure}[t]
  \centering
  \includegraphics[width=0.495\linewidth]{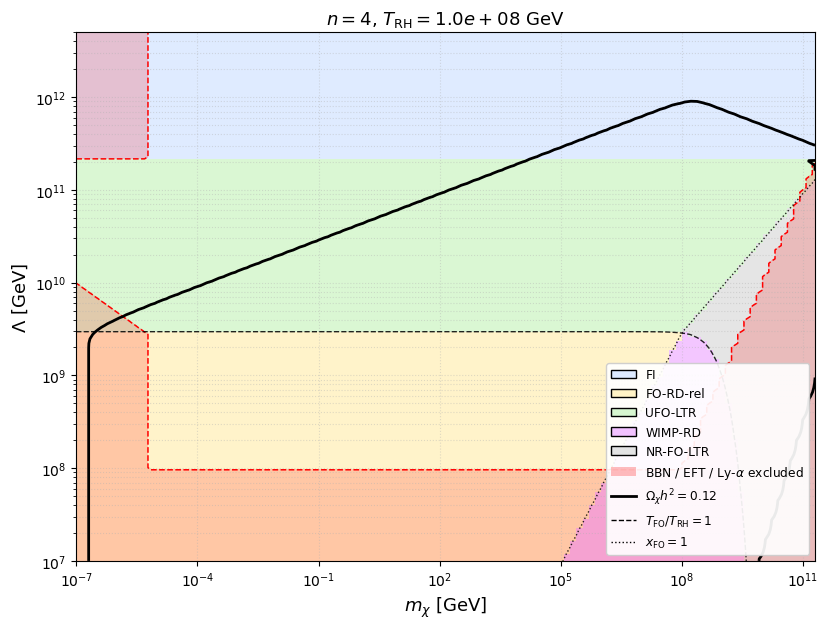}
  \includegraphics[width=0.495\linewidth]{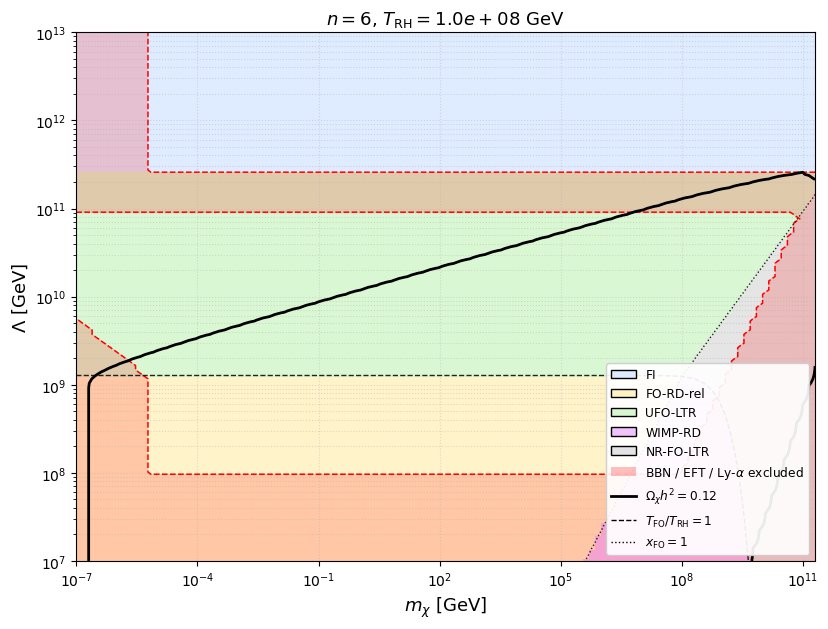}\\[4pt]
  \includegraphics[width=0.495\linewidth]{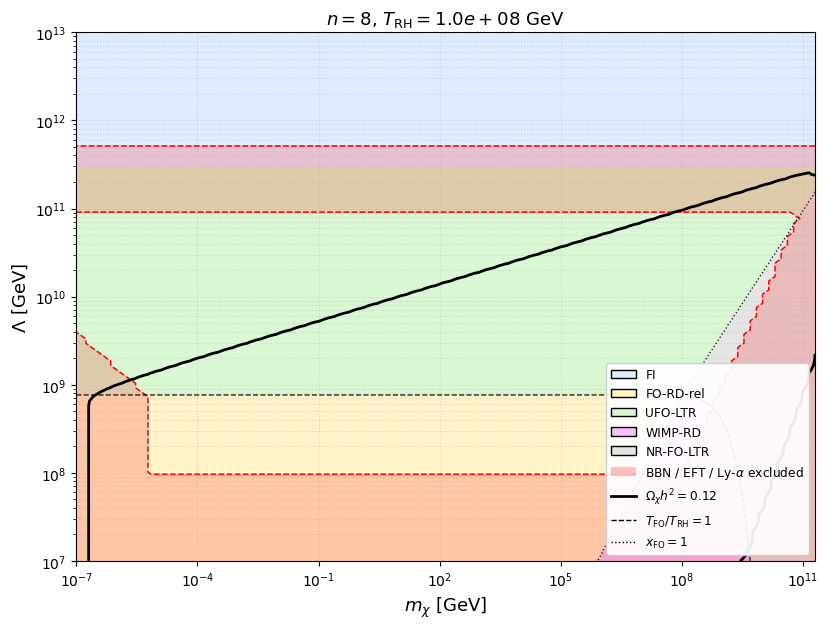}
  \caption{$(m_\chi,\Lambda)$ planes for $n=4,6,8$, with
  $\Trh=10^{8}$~GeV and $(\omega,\alpha)=(0,3/8)$. The background colours
  denote the numerical regime classification. The solid curve is the numerical
  $\Omega_\chi h^2=0.12$ contour. Shaded regions are excluded or flagged by the
  consistency conditions of Sec.~\ref{sec:bounds}.}
  \label{fig:Lam_vs_m}
\end{figure}

\paragraph{\boldmath The $(m_\chi,\Lambda)$ plane.}

Figure~\ref{fig:Lam_vs_m} shows the same physics at fixed $\Trh$. Since
\begin{equation}
  R_{\rm RH}\propto \Lambda^{-(n+2)},
\end{equation}
decreasing $\Lambda$ strengthens the interaction. The sequence from large to
small $\Lambda$ is therefore FI $\to$ LTR UFO/FO $\to$ RD freeze-out.

For the IR-dominated $n=4$ panel, Eq.~\eqref{eq:plotscaling_IR_UR} gives
\begin{equation}
  \Lambda_{\rm relic}\propto m_\chi^{1/(n+2)}
  \qquad
  (m_\chi\ll\Trh).
  \label{eq:plotslope_Lamm_IRUR}
\end{equation}
For $n=4$, this is $\Lambda\propto m_\chi^{1/6}$. Once the contour enters the
threshold-controlled region, Eq.~\eqref{eq:plotscaling_IR_thr} gives
\begin{equation}
  \Lambda_{\rm relic}
  \propto
  m_\chi^{\frac{n+1-n^*}{n+2}}.
  \label{eq:plotslope_Lamm_IRthr}
\end{equation}
For EMD this becomes
\begin{equation}
  \Lambda_{\rm relic}\propto m_\chi^{(n-5)/(n+2)}.
\end{equation}
Thus the $n=4$ contour changes from a shallow positive slope for
$m_\chi<\Trh$ to a negative slope for $m_\chi>\Trh$.

At $n=6$, the logarithmic factor is cut off by
$\max(\Trh,m_\chi)$ rather than by a new threshold power law. Since $\Trh$ is
fixed in this plane, the leading scaling is
\begin{equation}
  \Lambda_{\rm relic}
  \propto
  \left[
    m_\chi
    \ln\frac{T_{\rm in}}{\max(\Trh,m_\chi)}
  \right]^{1/8}.
\end{equation}
Thus crossing $m_\chi=\Trh$ produces at most mild logarithmic curvature. The
main turn occurs only when $m_\chi$ approaches $T_{\rm in}$, again typically
$T_{\rm FO}$ in the UFO branch.

For $n=8$, the UFO branch is UV dominated and follows
Eq.~\eqref{eq:plotscaling_UV} rather than the IR threshold law. In the EMD
benchmark,
\begin{equation}
  \Lambda_{\rm relic}^{\rm(UFO,UV)}
  \propto
  m_\chi^{(n-1)/[5(n+2)]},
\end{equation}
which is a shallow positive power. The bend in the contour occurs when
$m_\chi$ approaches the actual decoupling scale, not when it crosses
$\Trh$.

A useful contrast is that, for EMD with $n=4$, the UFO--FI transition does not
produce a visible slope change in the UR part of the contour: both pure FI and
deep IR-UFO are controlled near $\Trh$ and obey the same
$\Lambda\propto m_\chi^{1/6}$ scaling.


\begin{figure}[t]
  \centering
  \includegraphics[width=0.495\linewidth]{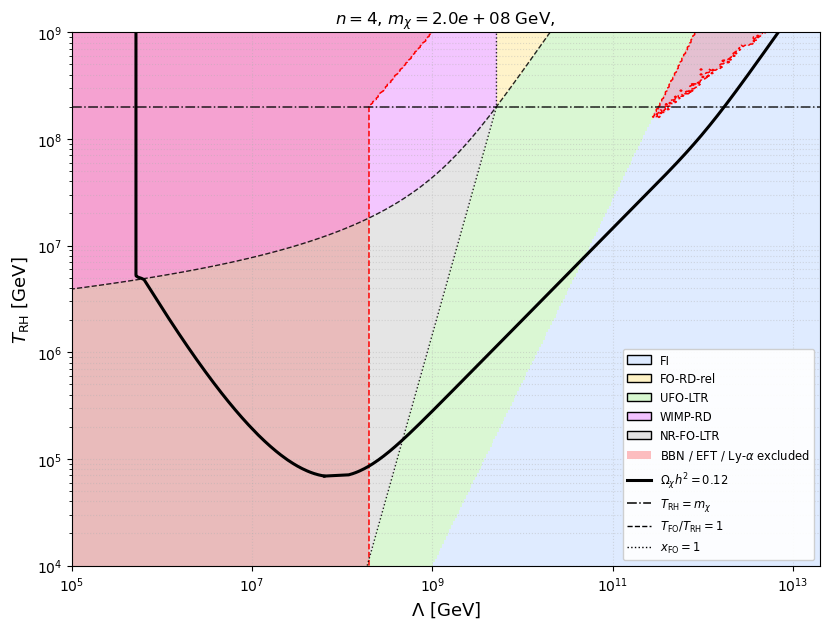}
  \includegraphics[width=0.495\linewidth]{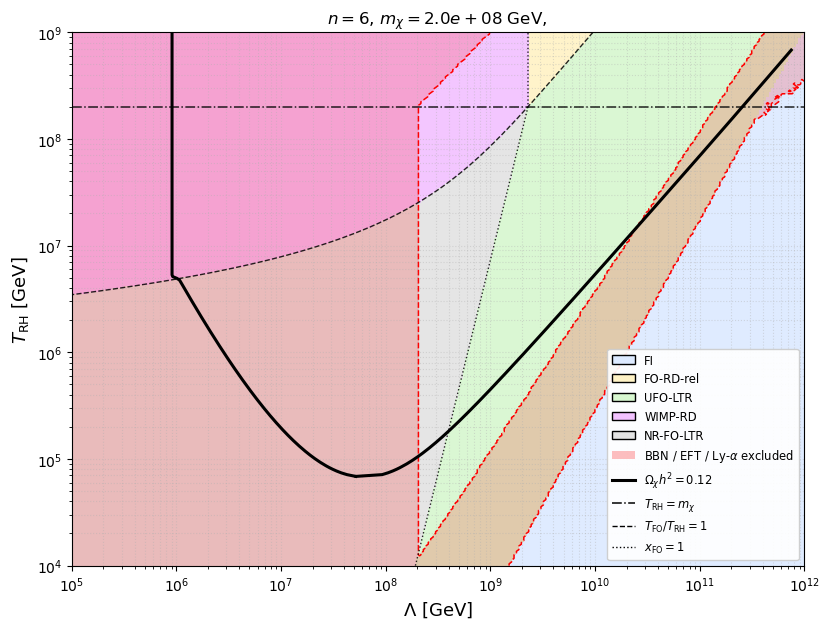}\\[4pt]
  \includegraphics[width=0.495\linewidth]{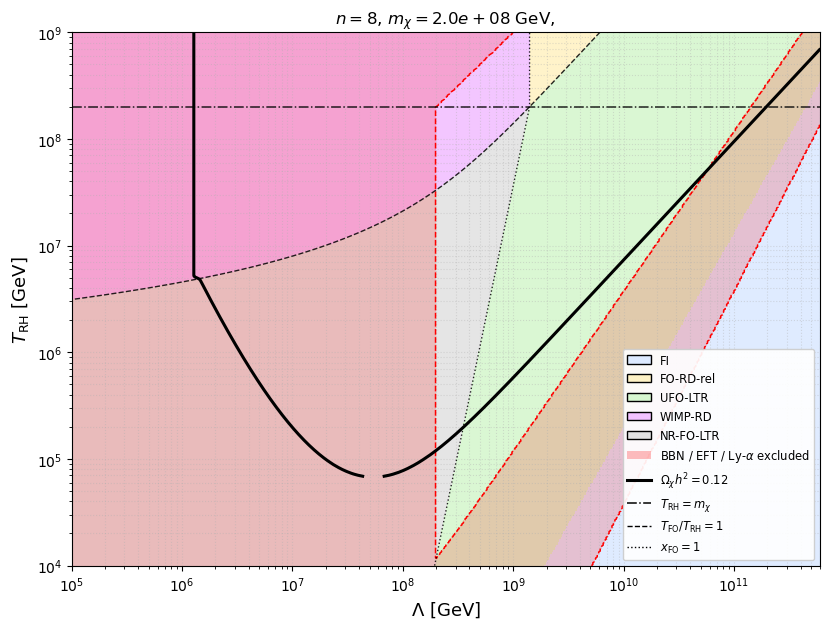}
  \caption{$(\Lambda,\Trh)$ planes for $n=4,6,8$, with fixed mass $m_\chi = 2 \times 10^8$ GeV and
   $(\omega,\alpha)=(0,3/8)$. The background colours
  denote the numerical regime classification. The solid curve is the numerical
  $\Omega_\chi h^2=0.12$ contour. Shaded regions are excluded or flagged by the
  consistency conditions of Sec.~\ref{sec:bounds}.}
  \label{fig:Trh_vs_Lam}
\end{figure}

\paragraph{\boldmath The $(\Lambda,\Trh)$ plane.}

The fixed-mass plane in Fig.~\ref{fig:Trh_vs_Lam} is useful for visualising the
transition between RD freeze-out and LTR-controlled production, but it is less
diagnostic of IR versus UV control than the two mass-dependent planes.

At small $\Lambda$, the interaction is strong and freeze-out occurs after
reheating. Since $m_\chi$ is fixed, the RD freeze-out relic density is
essentially independent of $\Trh$, producing an approximately vertical branch.
This branch includes both FO-RD-rel and WIMP-RD regions, depending on the value
of $x_{\rm FO}$.

Moving to larger $\Lambda$, freeze-out is pushed earlier and the contour passes
through an intermediate NR-FO-LTR region before reaching the UR LTR branches.
This is the softened bottom of the V-shaped structure. It should not be
interpreted as a separate asymptotic law; it is the semi-relativistic transition
between RD freeze-out, non-relativistic LTR freeze-out and the UR LTR regimes.
In the displayed scans this region is also where the EFT/uni\-tarity diagnostic
often becomes relevant.

For $n=4$, the LTR branch crosses the threshold line
$\Trh\simeq m_\chi$. In the UR IR-controlled part,
Eq.~\eqref{eq:plotscaling_IR_UR} gives
\begin{equation}
  \Lambda_{\rm relic}\propto \Trh^{5/6},
  \qquad
  (n=4,\ m_\chi\ll\Trh),
\end{equation}
whereas in the threshold-controlled region,
Eq.~\eqref{eq:plotscaling_IR_thr_benchmark} gives
\begin{equation}
  \Lambda_{\rm relic}\propto \Trh^{7/6},
  \qquad
  (n=4,\ \Trh\lesssim m_\chi).
\end{equation}
This is the origin of the visible change of slope as the contour leaves the
NR-FO-LTR transition and enters the IR-UFO/FI branch.

For $n=6$, the same comparison gives
\begin{equation}
  \Lambda_{\rm relic}\propto \Trh^{7/8}
\end{equation}
on both sides of the threshold, up to logarithmic corrections. Thus the edge of
the V is much smoother.

For $n=8$, the UV-UFO branch is controlled by $T_{\rm FO}$ rather than
$m_\chi\simeq\Trh$. Consequently, the contour does not react strongly to the
line $\Trh=m_\chi$. The main distortion instead comes from the approach
to the semi-relativistic regime and from the EFT/uni\-tarity diagnostic when
the relevant production scale approaches the cutoff.

\subsection{Beyond EMD: UV-dominated reheating histories and kination}
\label{subsec:beyond_emd_results}

We now compare the EMD benchmark with reheating histories in which the same
$n=4$ operator lies in a UV-dominated regime. The most useful plane for this
comparison is $(m_\chi,\Lambda)$ at fixed $\Trh$, because varying $\Lambda$
directly scans from RD freeze-out to LTR UFO and finally to FI.

\begin{figure}[t]
  \centering
  \includegraphics[width=0.495\linewidth]{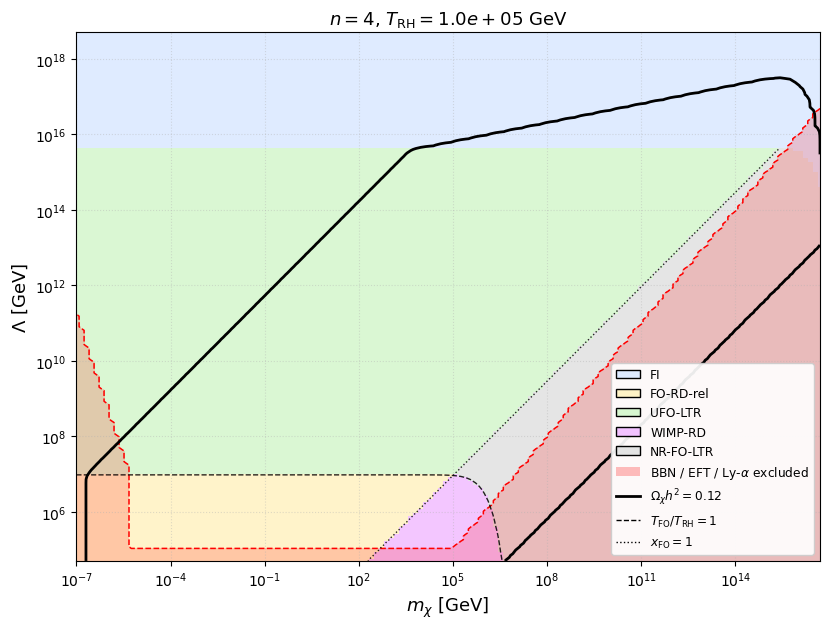}
  \includegraphics[width=0.495\linewidth]{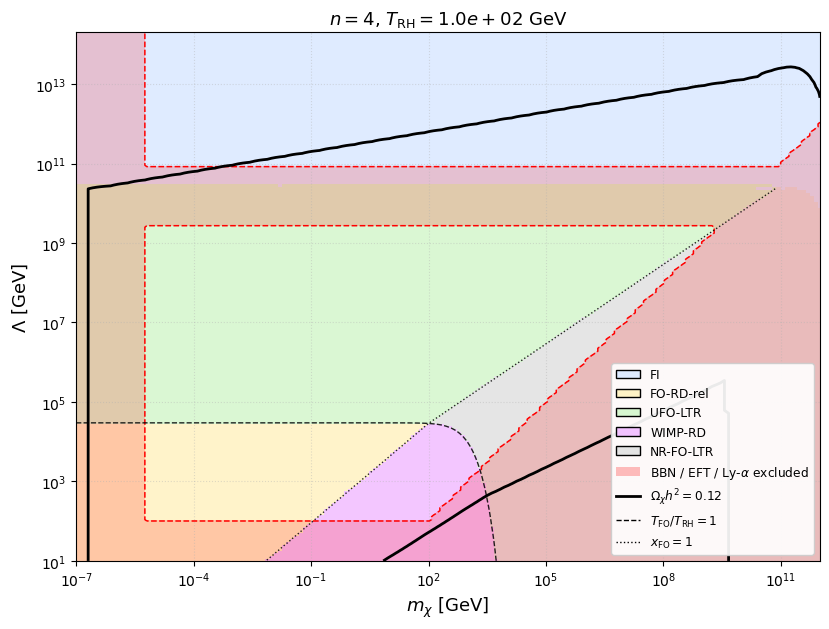}
  \caption{$(m_\chi,\Lambda)$ planes for $n=4$. Left: the motivated UV-dominated benchmark
  $(\omega,\alpha)=(1/3,3/4)$ with \(\Trh=10^5\,{\rm GeV}\). Right: kination,
  $(\omega,\alpha)=(1,1)$  with
  \(\Trh=10^2\,{\rm GeV}\). The background colours denote the numerical regime
  classification. The solid curve is the numerical $\Omega_\chi h^2=0.12$
  contour. Shaded regions are excluded or flagged by the consistency conditions
  of Sec.~\ref{sec:bounds}.}
  \label{fig:Lam_m_beyond_EMD}
\end{figure}

For the left panel of Fig.~\ref{fig:Lam_m_beyond_EMD}, we take
\begin{equation}
  (\omega,\alpha)=\left(\frac13,\frac34\right),
  \qquad
  \gamma_2=\frac83,
  \qquad
  n_c=-\frac13,
  \qquad
  n^*=\frac23,
  \qquad
  d=1.
\end{equation}
This benchmark can be motivated by a quartic monomial condensate with a
fermionic reheating channel. For $n=4$, one has $n>n^*$, so both UFO and FI are
UV dominated. However, the two branches are controlled by different physical
scales. In pure UV-FI, the yield is controlled by $T_I$, and
Eq.~\eqref{eq:plotscaling_UV_FI} gives
\begin{equation}
  \Lambda_{\rm relic}^{\rm(FI,UV)}
  \propto
  m_\chi^{1/6}.
\end{equation}
In the UV-UFO branch, the relic is controlled by $T_{\rm FO}$, and
Eq.~\eqref{eq:plotscaling_UV} gives
\begin{equation}
  \Lambda_{\rm relic}^{\rm(UFO,UV)}
  \propto
  m_\chi^{13/18}.
\end{equation}
The visible change of slope at the transition from the UV-FI region to the UV-UFO region is therefore a direct
signature that the UFO and FI branches are controlled by different UV scales.
This should be contrasted with the EMD $n=4$ panel in
Fig.~\ref{fig:Lam_vs_m}, where $n=4<n^*=6$ and both deep UFO and pure FI are IR
controlled, giving the same $\Lambda\propto m_\chi^{1/6}$ slope in the UR
region.

The right panel shows kination,
\begin{equation}
  (\omega,\alpha)=(1,1),
  \qquad
  \gamma_2=3,
  \qquad
  n_c=n^*=0,
  \qquad
  d=0.
\end{equation}
The limit $d=0$ is special. The entropy-dilution factor in the inherited UFO
yield is absent, so Eq.~\eqref{eq:plotscaling_UV} does not apply. Instead,
Eq.~\eqref{eq:kination_UFO_yield} shows that the deep-UFO yield remains of
order $Y_{\rm eq}$. In the deep relativistic UFO limit, the inherited yield remains
$Y_{\rm inh}\simeq Y_{\rm eq}$, so the relic density is fixed mainly by
$m_\chi$ rather than by a dilution-induced power of $\Lambda$. The transition to pure FI occurs
near the thermalisation boundary $R(T_I)\sim1$.

Together, these examples illustrate the main qualitative message of the
analysis. In entropy-producing reheating histories with $d>0$, UV-UFO can
generate relic-density contours with slopes that differ from pure FI. In
kination, where $d=0$, UFO is not diluted and instead resembles a relativistic
thermal relic.
\section{Conclusions}
\label{sec:conclusions}

We have developed a unified analytic description of ultra-relativistic freeze-out
and freeze-in during low-temperature reheating. The reheating background was
encoded by two parameters, $(\omega,\alpha)$, while the relativistic interaction
rate was parametrised by $\langle\sigma v\rangle=T^n/\Lambda^{n+2}$.
This setup captures a broad class of reheating histories and interaction
structures without committing to a specific mediator model.

The dynamics are organised by two critical indices. The first,
$n_c=\gamma_2-3$, determines whether a thermalised relativistic species can
freeze out during reheating or remains coupled until radiation domination. The
second, $n^*=\gamma_2-6+3/\alpha$, determines whether the FI-like production
integrals are IR dominated, logarithmic, or UV dominated. Together these
indices provide a compact phase diagram for FI, UFO-LTR, relativistic
post-reheating freeze-out, and non-relativistic freeze-out.

We derived analytic yields and relic-density targets in the different regimes,
including the inherited freeze-out abundance, entropy dilution, post-freeze-out
production, and the RD tail. In the ultra-relativistic regime this gives closed
expressions for $\Lambda_{\rm relic}$ as a function of $\Trh$ and
$m_\chi$. We also showed how threshold effects modify the contour slopes when
$m_\chi\gtrsim \Trh$, and how the logarithmic boundary $n=n^*$ differs
from genuinely IR-dominated cases.

The resulting relic-density contours depend strongly on the reheating history.
In matter-like reheating with $\alpha=3/8$, the $n=4$ branch is IR dominated,
so deep UFO and pure FI connect smoothly in the ultra-relativistic region. In
UV-dominated reheating histories, the UFO and FI branches are controlled by
different scales, $T_{\rm FO}$ and $T_I$, leading to visibly different contour
slopes. In kination, $d=0$ and the inherited UFO abundance is not diluted,
so the UFO branch resembles an undiluted relativistic relic. These examples
show that the relic-density target associated with a given interaction scale is
cosmology-dependent.

Our results provide the cosmology-side map needed to embed UFO production in
explicit particle-physics models. A UV completion specifies the Lorentz
structure, mediator mass and couplings, and therefore determines how the
effective scale $\Lambda$ maps onto collider, direct-detection or
indirect-detection observables. Combining such model-dependent probes with the
reheating phase diagram developed here would allow future studies to test not
only the dark matter interaction, but also the pre-BBN expansion history.

\appendix

\section{Details of analytic scalings}
\label{app:analytic_details}

This appendix contains the algebra behind the compact expressions used in
Sec.~\ref{sec:relic}. We keep the same assumptions as in the main text:
constant \(g_*\) and \(g_{*s}\), a single power-law interaction
\(\langle\sigma v\rangle=T^n/\Lambda^{n+2}\), and a monotonic LTR temperature
law \(T\propto a^{-\alpha}\).

\subsection{FI-like master integral}

Introducing
\begin{equation}
  x\equiv\frac{a}{\arh},
  \qquad
  x_{\rm in}\equiv\frac{a_{\rm in}}{\arh}
  =\left(\frac{\Trh}{T_{\rm in}}\right)^{1/\alpha},
\end{equation}
Eq.~\eqref{eq:J_a_scaling} gives
\begin{equation}
  N(a_{\rm in}\to\arh)
  =
  \mathcal J_{\rm RH}\arh
  \begin{cases}
    \dfrac{1-x_{\rm in}^{\alpha(n^*-n)}}{\alpha(n^*-n)},
    & n\neq n^*,
    \\[10pt]
    \ln\dfrac{1}{x_{\rm in}},
    & n=n^* .
  \end{cases}
  \label{eq:N_FI_master}
\end{equation}
Dividing by \(\mathcal S_{\rm RH}=s_{\rm RH}\arh^3\) and using
\begin{equation}
  \frac{\mathcal J_{\rm RH}}{s_{\rm RH}\arh^2}
  =
  \frac{\langle\sigma v\rangle_{\rm RH}n_{\rm eq}^2(\Trh)}
       {\Hrh s_{\rm RH}}
  =
  R_{\rm RH}Y_{\rm eq},
\end{equation}
one obtains
\begin{equation}
  Y_{\rm FI}(T_{\rm in}\to\Trh)
  =
  \begin{cases}
    \dfrac{Y_{\rm eq}}{\alpha(n^*-n)}R_{\rm RH}
    \left[1-x_{\rm in}^{\alpha(n^*-n)}\right],
    & n\neq n^*,
    \\[12pt]
    Y_{\rm eq}R_{\rm RH}\ln\dfrac{1}{x_{\rm in}},
    & n=n^* .
  \end{cases}
  \label{eq:YFI_exact}
\end{equation}
This is Eq.~\eqref{eq:YFI_master} after writing the result in terms of
temperatures.

\subsection{Finite-endpoint formulae and relic scales}

For \(n<n^*\), pure LTR freeze-in gives
\begin{equation}
  Y_0^{\rm(FI)}
  =
  \frac{Y_{\rm eq}}{\alpha(n^*-n)}R_{\rm RH}
  \left[1-
    \left(\frac{\Trh}{T_I}\right)^{n^*-n}
  \right].
  \label{eq:IR_FI_full}
\end{equation}
In the same regime, an LTR UFO abundance before adding the RD tail is
\begin{equation}
  Y_{\rm LTR}^{\rm(FO)}
  =
  Y_{\rm eq}\left(\frac{\Trh}{T_{\rm FO}}\right)^d
  +
  \frac{Y_{\rm eq}}{\alpha(n^*-n)}R_{\rm RH}
  \left[1-
    \left(\frac{\Trh}{T_{\rm FO}}\right)^{n^*-n}
  \right].
  \label{eq:IR_FO_full}
\end{equation}
For \(T_I\gg\Trh\) or \(T_{\rm FO}\gg\Trh\), the leading IR expression is
Eq.~\eqref{eq:IR_FI_final}. Setting it equal to \(Y_{\rm DM}\) gives
Eq.~\eqref{eq:Lambda_relic_FI}, or equivalently
\begin{equation}
  \Lambda_{\rm relic}^{\rm(IR)}
  \propto
  \left(m_\chi\Trh^{n+1}\right)^{1/(n+2)} .
  \label{eq:Lambda_FI_scaling}
\end{equation}

At the logarithmic point, the post-freeze-out piece is
\begin{equation}
  Y_{\rm post}
  =
  \frac{Y_{\rm eq}}{\alpha}R_{\rm RH}
  \ln\frac{T_{\rm FO}}{\Trh} .
  \label{eq:log_postFO}
\end{equation}
The same formula with \(T_{\rm FO}\to T_I\) gives pure logarithmic FI.

For \(n>n^*\), pure freeze-in is
\begin{equation}
  Y_0^{\rm(FI)}
  =
  \frac{Y_{\rm eq}}{\alpha(n-n^*)}R_{\rm RH}
  \left[
    \left(\frac{T_I}{\Trh}\right)^{n-n^*}-1
  \right].
  \label{eq:UV_FI_full}
\end{equation}
In the UV-dominated limit this reduces to Eq.~\eqref{eq:UV_FI_final}, and the
corresponding relic interaction scale is
\begin{equation}
  \Lambda_{\rm relic}^{\rm(FI,UV)}
  =
  \left[
    \frac{Y_{\rm eq}C_\Gamma M_P}{\alpha(n-n^*)Y_{\rm DM}}
    \sqrt{\frac{90}{\pi^2g_{\rm RH}}}
    \Trh^{n+1}
    \left(\frac{T_I}{\Trh}\right)^{n-n^*}
  \right]^{1/(n+2)} .
  \label{eq:Lambda_relic_UV_FI}
\end{equation}
Thus, at fixed \((\Trh,T_I)\),
\begin{equation}
  \Lambda_{\rm relic}^{\rm(FI,UV)}
  \propto m_\chi^{1/(n+2)} .
  \label{eq:Lambda_UV_FI_scaling}
\end{equation}

For UV-UFO, the post-freeze-out contribution can be written as
\begin{equation}
  Y_{\rm post}
  =
  \frac{Y_{\rm eq}}{\alpha(n-n^*)}R_{\rm RH}
  \left[
    \left(\frac{T_{\rm FO}}{\Trh}\right)^{n-n^*}-1
  \right].
  \label{eq:YIR_UV_exact}
\end{equation}
Using the freeze-out condition
\begin{equation}
  R(T_{\rm FO})
  =
  R_{\rm RH}
  \left(\frac{T_{\rm FO}}{\Trh}\right)^{\gamma_1-\gamma_2}
  =c_{\rm fo},
\end{equation}
and keeping the leading term for \(T_{\rm FO}\gg\Trh\), this becomes
\begin{equation}
  Y_{\rm post}
  \simeq
  \frac{c_{\rm fo}}{\alpha(n-n^*)}Y_{\rm eq}
  \left(\frac{\Trh}{T_{\rm FO}}\right)^d .
  \label{eq:UV_IR_in_terms_of_FO}
\end{equation}
Adding the inherited contribution gives Eq.~\eqref{eq:UV_with_subleading}. The
condition \(Y_0=Y_{\rm DM}\) then fixes
\begin{equation}
  \left(\frac{\Trh}{T_{\rm FO}}\right)^d
  =
  \frac{Y_{\rm DM}}{Y_{\rm eq}A_{\rm UV}} .
  \label{eq:TFO_from_Y_UV}
\end{equation}
The LTR freeze-out condition can also be written as
\begin{equation}
  \Lambda^{n+2}
  =
  \frac{C_\Gamma}{c_{\rm fo}}
  M_P
  \sqrt{\frac{90}{\pi^2g_{\rm RH}}}
  T_{\rm FO}^{\gamma_1-\gamma_2}
  \Trh^{\gamma_2-2} .
  \label{eq:Lambda_vs_TFO_exact}
\end{equation}
Combining the last two equations gives Eq.~\eqref{eq:Lambda_relic_UV}.

For \(\alpha=1\), \(d=0\), so the inherited relativistic abundance is not
diluted. In the deep-UFO limit,
\begin{equation}
  Y_{\rm LTR}^{\rm(FO)}
  \simeq
  Y_{\rm eq}
  \left[1+\frac{c_{\rm fo}}{\alpha(n-n^*)}\right],
  \qquad
  (\alpha=1,\; n>n^*) .
  \label{eq:kination_UFO_yield}
\end{equation}
Thus the kination UFO branch behaves like an undiluted relativistic thermal
relic rather than producing the diluted UV-UFO mass scaling.

\subsection{Threshold-controlled production}

When \(m_\chi\gtrsim\Trh\), an IR-dominated FI-like integral is cut off by the
mass threshold. Using the Maxwell--Boltzmann form
\begin{equation}
  n_{\rm eq}^{\rm(MB)}(T)
  \simeq
  g_\chi
  \left(\frac{m_\chi T}{2\pi}\right)^{3/2}
  e^{-m_\chi/T},
\end{equation}
one obtains
\begin{equation}
  Y_0^{\rm(FI-like)}
  =
  \frac{\mathcal C_{\rm thr}}{\alpha}
  \frac{M_P\Trh^{n^*+1}m_\chi^3}{\Lambda^{n+2}}
  \int_{\Trh}^{T_{\rm in}}dT\,
  T^{n-n^*-4}e^{-2m_\chi/T},
  \label{eq:Y_threshold_start}
\end{equation}
where, up to mild temperature dependence of the \(g\)-factors,
\begin{equation}
  \mathcal C_{\rm thr}
  =
  \frac{45\sqrt{90}}{16\pi^6}
  \frac{c_\sigma g_\chi^2}{g_{s,{\rm RH}}\sqrt{g_{\rm RH}}} .
\end{equation}
With \(x=m_\chi/T\),
\begin{equation}
  Y_0^{\rm(FI-like)}
  =
  \frac{\mathcal C_{\rm thr}}{\alpha}
  \frac{M_P\Trh^{n^*+1}}{\Lambda^{n+2}}
  m_\chi^{n-n^*}
  \int_{m_\chi/T_{\rm in}}^{m_\chi/\Trh}dx\,
  x^{n^*-n+2}e^{-2x} .
  \label{eq:Y_threshold_x}
\end{equation}
For \(n<n^*\) and \(\Trh\ll m_\chi\ll T_{\rm in}\), the remaining integral is
order unity and is dominated by \(x=\mathcal O(1)\). This gives
Eq.~\eqref{eq:Y_IR_threshold} and the abundance scaling
Eq.~\eqref{eq:Omega_IR_threshold}.

At \(n=n^*\), the relativistic part of the integral above threshold remains
logarithmic, but the lower limit is replaced by \(T\sim m_\chi\):
\begin{equation}
  Y_0^{\rm(log,thr)}
  \simeq
  \frac{Y_{\rm eq}}{\alpha}R_{\rm RH}
  \ln\frac{T_{\rm in}}{m_\chi}
  +\mathcal O(Y_{\rm eq}R_{\rm RH}) .
  \label{eq:Y_log_threshold_cutoff}
\end{equation}
Consequently the logarithmic branch does not acquire a new power-law slope when
\(m_\chi\) crosses \(\Trh\); the visible turnover occurs only as \(m_\chi\)
approaches the actual upper production scale \(T_{\rm in}\).

For the EMD benchmark, \((\omega,\alpha)=(0,3/8)\), one has \(n^*=6\). At
fixed \(\Lambda\), Eq.~\eqref{eq:Omega_IR_threshold} implies
\begin{equation}
  \Trh
  \propto
  m_\chi^{-(n+1-n^*)/(n^*+1)} ,
  \label{eq:Trh_m_threshold_scaling}
\end{equation}
while at fixed \(\Trh\),
\begin{equation}
  \Lambda_{\rm relic}
  \propto
  m_\chi^{(n+1-n^*)/(n+2)} .
  \label{eq:Lambda_m_threshold_scaling}
\end{equation}

\subsection{Genuine non-relativistic freeze-out}

Threshold-controlled FI-like production should not be confused with genuine
non-relativistic freeze-out. The latter begins only when the actual decoupling
temperature satisfies \(T_{\rm FO}\lesssim m_\chi\). Defining
\begin{equation}
  x_{\rm FO}\equiv\frac{m_\chi}{T_{\rm FO}},
\end{equation}
the radiation-dominated freeze-out condition gives parametrically
\begin{equation}
  x_{\rm FO}
  \simeq
  \ln\left[
    \frac{c'M_Pm_\chi^{n+1}}{\Lambda^{n+2}}
  \right]
  -
  \left(n-\frac12\right)\ln x_{\rm FO}
  +\cdots,
  \label{eq:xFO_NR}
\end{equation}
where \(c'\) is an order-one constant. The asymptotic yield can be estimated as
\begin{equation}
  Y_0^{\rm(NR)}
  \sim
  \frac{H(T_{\rm FO})}{s(T_{\rm FO})\langle\sigma v\rangle(T_{\rm FO})} .
  \label{eq:Y_NR_start}
\end{equation}
In RD this gives
\begin{equation}
  Y_0^{\rm(NR)}
  \sim
  \frac{\Lambda^{n+2}}{M_Pm_\chi^{n+1}}
  x_{\rm FO}^{n+1} .
  \label{eq:Y_NR_parametric}
\end{equation}
If non-relativistic freeze-out occurs during LTR, the same estimate should be
evaluated with the LTR Hubble rate and then multiplied by the entropy-dilution
factor in Eq.~\eqref{eq:entropy_ratio}.

\section{Microphysical realisations of \texorpdfstring{$(\omega,\alpha)$}{(omega,alpha)}}
\label{app:microphysics}

Here we briefly summarise some microphysical scenarios that can be mapped onto the $(\omega,\alpha)$ parametrisation used in the main text. Detailed derivations can be found in~\cite{Giudice:2000ex,Co:2020xaf,Garcia:2020wiy,Xu:2023lxw,Barman:2024mqo,Allahverdi:2020bys}.

\begin{itemize}
  \item \textbf{Matter-like reheating:} A quadratic potential $V(\phi)\propto\phi^2$ gives $\omega=0$. Perturbative decays into light species with constant decay width lead to $T\propto a^{-3/8}$, i.e.~$\alpha=3/8$.

  \item \textbf{Quartic potentials:} For $V(\phi)\propto\phi^4$, $\omega=1/3$; the condensate redshifts like radiation and reheating becomes less distinct from RD. Depending on the decay channel, $\alpha$ can interpolate between $1/2$ and $1$.

  \item \textbf{Constant-temperature reheating:} In certain inflaton--fermion couplings, Pauli blocking can keep the bath temperature approximately constant over a wide range of $a$, corresponding effectively to $\alpha\simeq 0$.

  \item \textbf{Annihilation-dominated reheating:} If the dominant energy transfer is via $\phi\phi\to$ SM annihilations, the temperature scaling can differ from the decay-dominated case, leading to different effective $\alpha$ at fixed $\omega$.

  \item \textbf{Kination:} For kinetic-energy domination, $\omega\simeq 1$ and the inflaton energy density redshifts faster than radiation. A simple effective description is $\alpha=1$, $T\propto a^{-1}$, with RD reached when $\rho_\phi\ll\rho_R$ even if $\phi$ does not fully decay.
\end{itemize}

\section{Entropy dilution and useful identities}
\label{app:entropy}

With $S(T)=s(T)a^3$ and $T\propto a^{-\alpha}$,
\begin{equation}
  S(T)\propto T^{3-3/\alpha}
  \quad\Rightarrow\quad
  \frac{S(T_{\rm FO})}{S(\Trh)}
  =\left(\frac{T_{\rm FO}}{\Trh}\right)^{3-3/\alpha},
\end{equation}
which is Eq.~\eqref{eq:entropy_ratio}. At \(n=n^*\), one has
\[
   \frac{\Gamma_{\rm eq}}{H}\propto T^{-(3-3/\alpha)} ,
\]
which implies
\begin{equation}
\frac{\Gamma_{\rm eq}(T_{\rm FO})}{H(T_{\rm FO})}
\frac{S(T_{\rm FO})}{S(\Trh)}
=
\left.\frac{\Gamma_{\rm eq}}{H}\right|_{\Trh} .
\label{eq:log_identity}
\end{equation}
This identity explains why the logarithmic terms in
Eqs.~\eqref{eq:YFI_master} and \eqref{eq:Ypost} can be written in terms of
\(R_{\rm RH}\).
\section*{Acknowledgments}
K.D. thanks Nicolás Bernal for discussions and suggestions on the topic and the
draft. OpenAI's ChatGPT (GPT-5.5) was used as an editorial aid for
proofreading, language polishing, and improving the clarity of author-written
text. The author is solely responsible for the scientific content of the
manuscript.

\bibliographystyle{JHEP}
\bibliography{biblio}
\end{document}